# A, B, C of three-qubit entanglement: three vectors to control it all

Dmitry B. Uskov [1,2*] and Paul M. Alsing [3]

[1]Department of Mathematics and Natural Sciences, Brescia University, Owensboro, Kentucky 42301, USA

[2]Department of Physics and Engineering Physics, Tulane University, New Orleans, Louisiana 70118, USA

[3]Air Force Research Laboratory, Information Directorate, Rome, New York 13441, USA

**Abstract.** In this paper we are focusing on entanglement control problem in a three-qubit system. We demonstrate that vector representation of entanglement, associated with $SO(6)$ representation of $SU(4)$ two-qubit group, can be used to solve various control problems analytically including (i) the transformation between a **W**-type states and **GHZ** state, and (ii) manipulating bipartite concurrences and three-tangle under a restricted access to only two qubits, and (iii) designing $USp(4)$-type quaternionic operations and quantum states.

*Corresponding author: dmitry.uskov@brescia.edu

## I. INTRODUCTION

The goal of this paper is to demonstrate how various single-qubit and two-qubit gates in a three-qubit system are represented geometrically as orthogonal rotations an approach developed in the paper [1] which to a large extent was inspired by [2-7]. First, we give a set of simple examples of state and entanglement manipulation, and then we describe the problem of entanglement boost showing that one can obtain analytical description of numerical optimization results using vector technique.

Since our work is using some Lie group theory (including $SU(2)$, $SO(3)$, $SU(4)$, $USp(4)$, $SL(2)$, $SO(3,C)$ and $SO(6)$ groups [8-10]) we first discuss some physical facts relevant to our work to set up a roadmap for further developments.

There are two physically consequential homomorphisms [11] between $SU$ and $SO$ Lie groups: low-dimensional $SU(2) \to SO(3)$ and higher-dimensional $SU(4) \to SO(6)$ [10]. In mathematical literature these are called accidental homomorphisms; algebraically they are linked to matrix representations of $Spin(N)$ groups – universal double covers of $SO(N)$ groups, i.e. a quotient group $Spin(N)/\mathbb{Z}_2$, $\mathbb{Z}_2 = \{\pm 1\}$ is isomorphic to $SO(N)$ [12,8]. Complexification of $SU$ group generates $SL$ group and this homomorphism takes more general form of $SL(2) \to SO(3,C)$, $SL(4) \to SO(6,\mathbb{C})$ [8,10]. There exists homomorphisms $SU(2,2) \to SO(4,2)$ between non-compact groups. However, implications of this mathematical feature of Lie groups have not yet been explored, at least in the quantum information context.

We start our discussion with $SU(2) \to SO(3)$ homomorphism which plays central role in setting up Bloch sphere description of spin ½ (or a qubit) state [13]. This example clearly illustrates that in addition to group homomorphism, 3D visualization of quantum states requires an additional mathematical feature - a map between $\mathbb{C}^2$ representation space of $SU(2)$ group and $\mathbb{R}^3$ representation space of $SO(3)$ group, i.e. a map from the space of quantum states to the space of real-valued three-dimensional vectors. Physically, the map between groups provides only a connection between $2 \times 2$ unitary matrices and $3 \times 3$ matrices describing three-dimensional rotations (we are discussing fundamental representations of both groups). A map between representation spaces grantees that quantum states are visualized in the form of three-dimensional vectors. Algebraically the map from Hilbert space $(x_1, x_2) \in \mathbb{C}^2$ to $\mathbb{R}^3$ space of real-valued vectors $(y_1, y_2, y_3)$ has the form $(y_1, y_2, y_3) = (\operatorname{Re} x_1 x_2^*, \operatorname{Im} x_1 x_2^*, |x_1|^2 - |x_2|^2)$. For physical (i.e. normalized) states we have $\operatorname{Re} x_1^2 + \operatorname{Im} x_1^2 + \operatorname{Re} x_2^2 + \operatorname{Im} x_2^2 = 1$ such that topologically $(x_1, x_2) \in S^3$. Since $y_1^2 + y_2^2 + y_3^2 = \left(|x_1|^2 + |x_2|^2\right)^2$ we also have $(y_1, y_2, y_3) \in S^2$. The map $(x_1, x_2) \to (y_1, y_2, y_3)$ is called first Hopf

fibration between spheres $S^2 \approx S^3/S^1$ [14]. Notice that the map between groups $SU(2) \to SO(3)$ differs even topologically: the group map is the map between simply connected $SU(2)$ manifold, which accidently also happens to be a sphere $S^3$, and $SO(3)$ manifold which is represented by a quotient space $S^3/\mathbb{Z}_2$. Technically, two opposite points of $S^3$ are considered as equivalent elements representing a single element of the quotient space $S^3/\mathbb{Z}_2$ (this space is isomorphic to real projective space $\mathbb{R}P^3$).

In physics we have so-called Bloch sphere [13]: density operators $\rho$ are expanded in $\sigma_{x,y,z}$ Pauli basis $\rho = 1/2\left(I + \sum_{n=1,2,3} m_n \sigma_n\right)$. The set of three expansion coefficients $m_{1,2,3}$ is called Bloch vector or Poincare sphere of photon polarizations [15]. Lie algebra space of $su(2)$ is three-dimensional and the adjoint representation of $SU(2)$ group is isomorphic to the fundamental representation $SO(3)$ [8-10,16]. For pure states $\rho = |\psi\rangle\langle\psi|$ and $m_1^2 + m_2^2 + m_3^2 = 1$, i.e. $\vec{m} \in S^2$.

Physical situation with $SU(4) \to SO(6)$ case is mathematically more complicated: there is no map from two-qubit $\mathbb{C}^4$ Hilbert space to $\mathbb{C}^6$ space, consistent with the group map. Notice that adjoint representation of $SU(4)$ group is 15-dimensional while fundamental representation of $SO(6)$ is 6-dimensional such that direct generalization of spin-1/2 Bloch sphere design does not work in this case. Yet, in the context of quantum information theory $SU(4)$ group is of fundamental importance because it describes interaction between two qubits which is at the core of quantum information processing [17].

To use mathematical potential of $SU(4) \to SO(6)$ homomorphism for quantum mechanical applications one should introduce more sophisticated algebraic constructions which ultimately lead to restriction of this tool to geometric description of entanglement in a few-qubit quantum systems rather than geometry of quantum states per se [1, 2, 5, 6]. The main idea can be described as follows. Consider a set of pairs of two-qubit quantum states. Each pair spans $\mathbb{C}^2$ complex two-dimensional subspace of $\mathbb{C}^4$. A collection of all $\mathbb{C}^2 \subset \mathbb{C}^4$ subspaces is called Grassmannian $\mathrm{Gr}_2(\mathbb{C}^4)$, which is a compact complex $4$-dimensional manifold (see, for example, [18]). $SU(4)$ group acts on elements of $\mathrm{Gr}_2(\mathbb{C}^4)$ by acting on pairs of non-parallel vectors from $\mathbb{C}^4$. This action does not depend on the choice of a pair of vectors representing specific subspace. While the $\mathrm{Gr}_2(\mathbb{C}^4)$ is not a linear space, the space of homogeneous coordinates on this space, called Plücker coordinates [19], is a special subspace of $\mathbb{C}^6$. The $SU(4)$ action on this space takes the form of six-dimensional representation of $SU(4)$ group which is isomorphic to the fundamental representation of $SO(6)$ group [20].

Following the route outlined above, we found that $SU(4) \to SO(6)$ homomorphism provides natural mathematical tool for geometric analysis of i) partition independent description of three-qubit entanglement and, more importantly, ii) geometric tool for designing control operations involving qubit-qubit coupling [1].

## II. VECTORS **A**, **B** AND **C** AS INVARIANTS OF TWO-QUBIT LOCAL $SL(2)$ TRANSFORMATIONS

In this section we consider $SL(2)$ invariants, as well as $SU(2)$ invariants, since the former are relevant for quantum communication protocols and measures of entanglement which can involve Local Operations and Classical Communications (LOCC) (for details see review [21] page 903). The $SU(2)$ group is a subgroup of $SL(2)$ therefore any invariant of $SL(2)$ is also an invariant of $SU(2)$. However, the converse is not true. For example, the norm of a state is not an invariant of $SL(2)$ action.

To avoid possible confusion concerning Lie group terminology we note that there are two versions of $SL(2)$ group: $SL(2,\mathbb{R})$ and $SL(2,\mathbb{C})$. In quantum mechanics we are dealing, by default, with $SL(2,\mathbb{C})$ group, i.e. complex invertible matrices with unit determinant.

Consider a system of thee qubits, labeled as $a$, $b$ and $c$, described by the wave function

$$|\psi\rangle = \sum_{i,j,k=0,1} c_{ijk} |i,j,k\rangle \qquad (1)$$

A simple method of constructing $SU(2)^{(a,b,c)}$ and $SL(2)^{(a,b,c)}$ local invariants is to consider arrangements of coefficients $c_{ijk}$ in the form of $4\times 2$ matrices corresponding to three bipartite arrangements: $a(bc)$, $b(ca)$ and $c(ab)$

$$\begin{pmatrix} c_{0,0,0} & c_{1,0,0} \\ c_{0,0,1} & c_{1,0,1} \\ c_{0,1,0} & c_{1,1,0} \\ c_{0,1,1} & c_{1,1,1} \end{pmatrix}, \begin{pmatrix} c_{0,0,0} & c_{0,1,0} \\ c_{1,0,0} & c_{1,1,0} \\ c_{0,0,1} & c_{0,1,1} \\ c_{1,0,1} & c_{1,1,1} \end{pmatrix}, \begin{pmatrix} c_{0,0,0} & c_{0,0,1} \\ c_{0,1,0} & c_{0,1,1} \\ c_{1,0,0} & c_{1,0,1} \\ c_{1,1,0} & c_{1,1,1} \end{pmatrix} \qquad (2a,b,c)$$

Notice that matrices (2b) and (2c) are obtained from matrix (2a) by permutations of index values $(i,j,k) \to (k,i,j)$ and $(i,j,k) \to (j,k,i)$ correspondingly. Local action on qubits $a$, $b$ and $c$ have the form of right multiplication by $SL(2)$ $2\times 2$ matrices, acting on the first, second and third matrix correspondingly. Since determinants of $SL(2)$ matrices are equal to 1, subdeterminants of matrices (2a,b,c) do not change under $SL(2)^{(a,b,c)}$ operations acting on corresponding qubits [5]. Formally, a set of six subdeterminants of a $4\times 2$ matrix define Plücker coordinates on $Gr_2(\mathbb{C}^4)$ manifold.

There are overall 18 subdeterminants: six subdeterminants for each arrangement in equation (2a,b,c). All of these 18 subdeterminants are quadratic polynomials in coefficients $c_{ijk}$. However, the space spanned by these polynomials is only 9-dimensional: there are nine linear dependences between these polynomials. Application of Lie group representation theory allows to construct partition-symmetric basis of nine linearly independent polynomials which form a complete basis in this space. These polynomials are naturally subdivided into three groups associated with partitions $a(bc)$, $b(ca)$ and $c(ab)$. We call these polynomial triples as 3-dimensional complex-valued vectors **A**, **B** and **C** [1].

For partition $a(bc)$ components of vector **A** are defined as

$$\begin{aligned} A_1 &= -i(c_{000}c_{011} - c_{001}c_{010} + c_{110}c_{101} - c_{111}c_{100}) \\ A_2 &= (c_{000}c_{011} - c_{001}c_{010} + c_{100}c_{111} - c_{101}c_{110}) \\ A_3 &= i(c_{000}c_{111} - c_{001}c_{110} + c_{100}c_{011} - c_{101}c_{010}) \end{aligned} \qquad (3a)$$

Vectors **B** and **C** are obtained from equation (3a) by permutations of index values $(i,j,k) \to (k,i,j)$ and $(i,j,k) \to (j,k,i)$ correspondingly.

$$\begin{aligned} B_1 &= -i(c_{000}c_{101} - c_{100}c_{001} + c_{011}c_{110} - c_{111}c_{010}) \\ B_2 &= (c_{000}c_{101} - c_{100}c_{001} + c_{010}c_{111} - c_{110}c_{011}) \\ B_3 &= i(c_{000}c_{111} - c_{100}c_{011} + c_{010}c_{101} - c_{110}c_{001}) \end{aligned} \qquad (3b)$$

$$\begin{aligned} C_1 &= -i(c_{000}c_{110} - c_{010}c_{100} + c_{101}c_{011} - c_{111}c_{001}) \\ C_2 &= (c_{000}c_{110} - c_{010}c_{100} + c_{001}c_{111} - c_{011}c_{101}) \\ C_3 &= i(c_{000}c_{111} - c_{010}c_{101} + c_{001}c_{110} - c_{011}c_{100}) \end{aligned} \qquad (3c)$$

It is easy to verify that vector **A** is a linear combination of subdeterminants of matrix (2b), and at the same time it is a combination of subdeterminant of matrix (2c). Similarly, vector **B** is a combination of subdeterminants of (2a), (2c) and vector **C** is a combination of subdeterminants (2a), (2b). Therefore each of **A**, **B**, **C** vectors represents triples of two-qubit $SL(2)$ invariants: **A**, **B** and **C** do not change under $SL(2)^{(b,c)}$, $SL(2)^{(a,c)}$ and $SL(2)^{(a,b)}$ action, correspondingly. Notice that canonical Plücker variables, defined as subdeterminants of (2a,b,c), are single-qubit invariants [5]. As we mentioned above, $SU(2)$ is a subgroup of $SL(2)$ and polynomials (3a,b,c) are also $SU(2)$ invariants. As we show in sections 4 and 7, these polynomial triples provide natural description of entanglement parameters in a three-qubit system as well as geometric description of entanglement evolution under two-qubit interaction, which is instrumental in analyzing and designing various three-qubit gates and transformations.

## III. BLOCH-TYPE SO(3) EVOLUTION OF VECTORS **A**, **B**, **C**

As discussed above, vectors **A**, **B**, **C** do not change under local action on qubits ($b$, $c$), ($c$, $a$) and ($b$, $a$) correspondingly. **A**, **B**, **C** vectors follow Bloch-type evolution under $SU(2)$ action on $a$, $b$, $c$ qubits correspondingly. Consider local transformations $\exp(-it\mathbf{H}^{local})$ generated by the Hamiltonian

$$\mathbf{H}^{local} = \frac{1}{2}\sum_{n=1}^{3}\left(\beta_n^{(a)}\sigma_n^{(a)} + \beta_n^{(b)}\sigma_n^{(b)} + \beta_n^{(c)}\sigma_n^{(c)}\right), \tag{4}$$

where $\sigma_{1,2,3}^{(a,b,c)}$ are $su(2)^{(a,b,c)}$ generators of local transformations on qubits $a$, $b$, $c$ (e.g. $\sigma_{1,2,3}^{(a)} = \sigma_{x,y,z}\otimes\mathbf{I}\otimes\mathbf{I}$, where $\sigma_{x,y,z}$ are standard Pauli matrices).

Evolution equations for **A**, **B**, **C** follow directly from Schrödinger equation for coefficients $c_{ijk}$ of the wave function (1). All three vectors evolve independently, i.e. vector **A** is not coupled to vectors **B** and **C**, etc.

$$d/dt\,\mathbf{A},\mathbf{B},\mathbf{C} = \left(\beta_1^{(a,b,c)}l_{3,2} - \beta_2^{(a,b,c)}l_{3,1} + \beta_3^{(a,b,c)}l_{2,1}\right)\mathbf{A},\mathbf{B},\mathbf{C} \tag{5}$$

where so(3) generators are

$$l_{2,1} = \begin{pmatrix} 0 & 1 & 0 \\ -1 & 0 & 0 \\ 0 & 0 & 0 \end{pmatrix}, l_{3,1} = \begin{pmatrix} 0 & 0 & 1 \\ 0 & 0 & 0 \\ -1 & 0 & 0 \end{pmatrix}, l_{3,2} = \begin{pmatrix} 0 & 0 & 0 \\ 0 & 0 & 1 \\ 0 & -1 & 0 \end{pmatrix} \tag{6a}$$

Relation between $SU(2)$ and $SO(3)$ forms is constructed as follows: for an SU(2) operator

$$\mathbf{U}_{su} = \exp\left(1/2\sum_{n=1}^{3}\theta_n i\sigma_n\right) \tag{6b}$$

the corresponding SO(3) rotation is

$$\mathbf{Y}_{so} = \exp\left(\theta_1 l_{3,2} - \theta_2 l_{3,1} + \theta_3 l_{2,1}\right). \tag{6c}$$

As Lie algebras $su(2)$ and $so(3)$ are isomorphic. The map

$$i/2\sigma_1 \to l_{3,2},\ i/2\sigma_2 \to -l_{3,1},\ i/2\sigma_3 \to l_{2,1} \tag{7}$$

generates identical commutation relations for both algebras. For example, $[i/2\sigma_1, -i/2\sigma_2] = i/2\sigma_3$ corresponds to $[l_{3,2}, l_{3,1}] = l_{2,1}$ etc. However, at the level of Lie groups we have a two-to-one map. Notice that while $\exp(\pi i\sigma_n) = -\mathbf{I}$, the corresponding SO(3) rotations $\exp(2\pi l_{n,m}) = \mathbf{I}$. As a result, $SU(2)$ operators different only by factor $(-\mathbf{I})$ will have identical $SO(3)$ image, i.e. $SU(2)$ is a double cover of $SO(3)$.

Equation (5) also can be written down in the form of a vector cross product (see for example [22])

$$d/dt\,\mathbf{A} = \boldsymbol{\beta}^{(a)} \times \mathbf{A}, \quad d/dt\,\mathbf{B} = \boldsymbol{\beta}^{(b)} \times \mathbf{B}, \quad d/dt\,\mathbf{C} = \boldsymbol{\beta}^{(c)} \times \mathbf{C} \qquad (8)$$

Notice that $SL(2)^{(a,b,c)}$ transformations are generated by allowing coefficients $\beta_{1,2,3}^{(a,b,c)}$ in equations (4,5) to take complex values [10]. This will also result in complex-valued parameters $\theta_{1,2,3}$ in equations (6b,c).

The reason why equations (5) have such a simple form is related to the following property of $SU(4) \to SO(6)$ homomorphism: local $SU(2)$ transformations are embedded in $SU(4)$ group as tensor product subgroup $SU(2) \otimes SU(2)$, while $SO(3)$ rotations are represented in $SO(6)$ as block-diagonal direct product $SO(3) \times SO(3)$ subgroup. Therefore local transformations of vectors $\mathbf{A}$, $\mathbf{B}$, $\mathbf{C}$ are algebraically decoupled for corresponding qubits.

## IV. THREE-QUBIT LOCAL INVARIANTS AND THREE-TANGLE

As we see, each three-component vector $\mathbf{A}$, $\mathbf{B}$, $\mathbf{C}$ represents three two-qubit local invariants. If we would like to construct three-qubit invariants we need to use an additional invariant specific to SO(3) group action, equation (5). As a classical isometry group, $SO(3)$ has vector dot product as invariant bilinear form. Since $SO(3)$ action does not change distance between point in $\mathbb{R}^3$ space, it also does not change dot product between vectors. Thus we have $\mathbf{A}^2$, $\mathbf{B}^2$ and $\mathbf{C}^2$ as three-qubit $SU(2)^{(a,b,c)}$ local invariants since any $SU(2)$ action on the wave function takes the form of corresponding $SO(3)$ action on $\mathbf{A}$ or $\mathbf{B}$ or $\mathbf{C}$ (for convenience, we denote the dot product $\mathbf{A} \cdot \mathbf{B}$ of vectors $\mathbf{A}$ and $\mathbf{B}$ by their juxtaposition $\mathbf{AB}$, and square of a vector means conventional dot product of a vector by itself).

It may not be obvious, but vector dot products will be also invariant relative to a larger class of local transformations generated by $SL(2,\mathbb{C})$ action. This group appears in the context of single qubit operations involving measurements as we indicated above. Explanation of this fact is provided by the following observation. $SL(2,\mathbb{C})$ group is isomorphic to complexification $SU(2)_\mathbb{C}$ of standard $SU(2)_\mathbb{R}$ group, generated as $\exp\left(\sum_{n=1}^{3} \xi_n i\sigma_n\right)$, $\xi_n \in \mathbb{C}$. Homomorphism $SU(2)_\mathbb{C} \to SO(3)_\mathbb{C}$ takes the form $SU(2) \cong SL(2,\mathbb{C}) \to SO(3)_\mathbb{C}$. As a result, $SL(2,\mathbb{C})$ action on the wave function is represented as $SO(3)_\mathbb{C}$ action on $\mathbf{A}$, $\mathbf{B}$ or $\mathbf{C}$ vectors. Both $SO(3)_\mathbb{R}$, corresponding to $SU(2)_\mathbb{R}$, and $SO(3)_\mathbb{C}$, corresponding to $SL(2,\mathbb{C})$, are represented by matrices satisfying equation $\mathbf{M}^T = \mathbf{M}^{-1}$, which guarantees dot product invariance relative to both types of action.

One may expect that for an arbitrary three qubit state numerical values of $\mathbf{A}^2, \mathbf{B}^2, \mathbf{C}^2$ will be different since the wave function does not have to be symmetric relative to qubit permutations. However, Plücker relation [19, 23] for subdeterminants guarantees that all three values are identical even if the wave function is not qubit-permutation symmetric, i.e.

$$\mathbf{A}^2 = \mathbf{B}^2 = \mathbf{C}^2. \qquad (9)$$

One expects that three-tangle (equations (22) and (25) of [24]), which is the only three-qubit quadratic $SL(2)^{(a,b,c)}$ invariant, is related to $\mathbf{A}^2, \mathbf{B}^2, \mathbf{C}^2$. Indeed, by direct inspection one finds that

$$\tau_{abc} = 4|\mathbf{B}^2| = 4|\mathbf{C}^2| = 4|\mathbf{A}^2| \qquad (10)$$

Notice that phase of $\mathbf{A}^2, \mathbf{B}^2, \mathbf{C}^2$, which is a well-defined quantity due to relation (9), is an additional unconventional $SL(2)^{(a,b,c)}$ and $SU(2)^{(a,b,c)}$ three-qubit invariant which is commonly overlooked since it is equivalent to a global phase factor (this invariant is absent in conventional definition of three-tangle in equation (10)).

We would like to stress that equation (10) for three-tangle is explicitly partition independent (i.e. qubit-permutation symmetric) due to Plücker relation [19] for matrix subdeterminants, equation (9) [23]. At the same time, the authors of seminal paper [24] make the following intriguing comment concerning partition independence of their equation (25) for three-tangle: "This form does not immediately reveal the invariance of $\tau_{abc}$ under permutations of the qubits, but the invariance is there nonetheless." And, indeed, algebraic transformations of this equation by index manipulations, for example, do not reveal any simple route to verify partition invariance of this equation. This is due to the fact that Plücker identity (9) is in the root of any explicit derivation of partition independence of this equation, and demonstration of this property will implicitly, or explicitly as in our case, involve Plücker identity for subdeterminants.

Second set of three-qubit $SU(2)^{(a,b,c)}$ invariants are two-tangles, or Wootters concurrences [25]. Since $SO(3)$ group is a subgroup of $SU(3)$ we have Hermitian product as an additional invariant such that $\mathbf{AA}^*$, $\mathbf{BB}^*$ and $\mathbf{CC}^*$ are also three-qubit invariants. We have two-tangles expressed as combinations of both invariants

$$\tau_{(bc)} = 2\left(\mathbf{AA}^* - |\mathbf{A}^2|\right), \quad \tau_{(ac)} = 2\left(\mathbf{BB}^* - |\mathbf{B}^2|\right), \quad \tau_{(ab)} = 2\left(\mathbf{CC}^* - |\mathbf{C}^2|\right) \tag{11a,b,c}$$

Notice that while dot product is $SL(2)$-invariant, general action of $SL(2)$ does not preserve Hermitian product and two tangles are not invariant relative to $SL(2)^{(a,b,c)}$ transformations.

## V. AN EXAMPLES OF LOCAL TRANSFORMATION: HADAMARD GATE

The Hadamard gate [17] acting, for example, on qubit $a$ of a three-qubit system has the following form

$$\mathbf{H}^{(a)} = e^{-i\pi/2} e^{i\pi/2\sigma_3^{(a)}} e^{i\pi/4\sigma_2^{(a)}(a)} \otimes \mathbf{I}^{(b)} \otimes \mathbf{I}^{(c)} \equiv \frac{1}{\sqrt{2}}\begin{pmatrix} 1 & 1 \\ 1 & -1 \end{pmatrix}^{(a)} \otimes \mathbf{I}^{(b)} \otimes \mathbf{I}^{(c)} \tag{12}$$

The action on vector $\mathbf{A}$, according to equations (7) is a sequence of $-\pi/2$ rotation in the $3-1$ coordinate plane, and $\pi$ rotation in the $2-1$ plane correspondingly

$$\exp(\pi l_{21})\exp(-\pi/2 l_{3,1}) = \begin{pmatrix} -1 & 0 & 0 \\ 0 & -1 & 0 \\ 0 & 0 & 1 \end{pmatrix}\begin{pmatrix} 0 & 0 & -1 \\ 0 & 1 & 0 \\ 1 & 0 & 0 \end{pmatrix} = \begin{pmatrix} 0 & 0 & 1 \\ 0 & -1 & 0 \\ 1 & 0 & 0 \end{pmatrix}, \tag{13a}$$

plus multiplication by a global phase factor $\exp(-i\pi/2)^2 = -1$. As a result, we have transformation

$$(A_1, A_2, A_3) \to (-A_3, A_2, A_1) \to (A_3, -A_2, A_1) \xrightarrow{\times(-1)} (-A_3, A_2, -A_1). \tag{13c}$$

Multiplication by a global phase factor $\exp(-i\pi/2)$ in equation (12) is not an element of $SU(2)$ group and multiplication of vector $\mathbf{A}$ by the corresponding factor $-1$ is not in $SO(3)$ group. This is an inversion operation which is an element of $O(3)$ group [8].

A local operator $\exp(i\vartheta/2\sigma_3^{(a)})$, for example, generates rotation $\exp(\vartheta l_{21})$ in the $2-1$ plane by the angle $\vartheta$:

$$\begin{pmatrix} A_1 \\ A_2 \\ A_3 \end{pmatrix} \to \begin{pmatrix} \cos\vartheta A_1 + \sin\vartheta A_2 \\ \cos\vartheta A_2 - \sin\vartheta A_1 \\ A_3 \end{pmatrix} \tag{14}$$

Notice that transformations (13c) and (14) do not change invariants (10) and (11a,b,c). Transformations (14) is an important part of a scheme for generating an arbitrary coupling transformation by concatenating two CNOT gates and a set of local transformations.

## VI. QUBIT-QUBIT COUPLING

For convenience, in this section three bipartite arrangements $a(bc)$, $b(ca)$ and $c(ab)$ will be referred to as 1, 2 and 3. Consider, for example, coupling Hamiltonian between qubits $a$ and $b$, associated with partition 3, corresponding to matrix arrangement (3c).

$$\mathbf{H}^{(ab)} = \frac{1}{2} \sum_{m,n=1}^{3} \gamma_{nm}^{(ab)} \sigma_{nm}^{(ab)} \tag{15}$$

Here we use the following short notations $\sigma_{nm}^{(ab)} \equiv \sigma_n^{(a)} \otimes \sigma_m^{(b)} \otimes \mathbf{I}^{(c)}$.

In general, any SU(4) transformations can be factored into a product of local transformations and a coupling operator. This form is usually called Global Cartan decomposition or Polar Decomposition (see [8] page 433). $\mathbf{U}^{(ab)} = \mathbf{U}_{coupling}^{(ab)} \mathbf{U}_{local}^{(a)} \mathbf{U}_{local}^{(b)}$. Local operators $\mathbf{U}_{local}^{(a,b)}$ correspond to $SO(3)^{(a)} \otimes SO(3)^{(b)}$ transformations $\mathbf{A}, \mathbf{B} \to \exp\left(\beta_1^{(a,b)} l_{3,2} - \beta_2^{(a,b)} l_{3,1} + \beta_3^{(a,b)} l_{2,1}\right) \mathbf{A}, \mathbf{B}$, equation (6c). The coupling operator $\mathbf{U}_{coupling}^{(ab)}$ contains complete set of qubit-qubit interaction terms

$$\mathbf{U}_{coupling}^{(ab)} = \exp\left[1/2 \sum_{n,m=1}^{3} \theta_{nm} i\sigma_{nm}^{(ab)}\right] \tag{16}$$

In order to formulate how coupling operations act on vectors we have to set up a rule for mapping $i\sigma_{nm}^{(ab)}$ generators of su(4) algebra onto corresponding so(6) generators describing orthogonal rotations of 6-dimensional vector. It is convenient to arrange vectors $\mathbf{A}$ and $\mathbf{B}$ in the form of a single 6-dimensional vector with components $\mathbf{Q}^{(3)} = (A_1, A_2, A_3, -iB_1, -iB_2, -iB_3) \equiv (\mathbf{A}, \tilde{\mathbf{B}})$, where entries 4–6 are associated with vector $\tilde{\mathbf{B}} = -i\mathbf{B}$. In this case SO(6) transformations have a canonical form of real-valued matrices generated by Lie algebra of real-valued antisymmetric so(6) matrices. For partition (1) and (2) we have $\mathbf{Q}^{(1)} = (\mathbf{B}, -i\mathbf{C}) \equiv (\mathbf{B}, \tilde{\mathbf{C}})$ and $\mathbf{Q}^{(2)} = (\mathbf{C}, -i\mathbf{A}) \equiv (\mathbf{C}, \tilde{\mathbf{A}})$ correspondingly.

A map between coupling su(4) and so(6) generators has the form (see Appendix for mathematical details)

$$i/2 \sigma_{nm}^{(ab)} \to \Lambda_{n,m}^{(ab)}, \quad n,m=1,2,3 \tag{17}$$

where $\Lambda_{n,m}^{(ab)}$ is a set of nine $6\times 6$ antisymmetric matrices with matrix elements

$$\left[\Lambda_{n,m}^{(ab)}\right]_{i,j} = -\delta_{i,n}\delta_{j,m+3} + \delta_{j,n}\delta_{i,m+3}, \quad i,j=1,2,3,4,5,6 \tag{18}$$

Notice that $\Lambda_{n,m}^{(ab)}$ are coupling $n^{th}$ component of $\mathbf{A}$ and $m^{th}$ component of $\tilde{\mathbf{B}}$, which corresponds to coupling of $n^{th}$ and $(m+3)^{th}$ components of 6-dimensional vector $\mathbf{Q}^{(3)} = (A_1, A_2, A_3, \tilde{B}_1, \tilde{B}_2, \tilde{B}_3)$. For example, a linear combination of generators $i\sigma_{11}^{(ab)}, i\sigma_{22}^{(ab)}, i\sigma_{33}^{(ab)}$ (i.e. maximal Abelian subalgebra of $i\sigma_{nm}^{(ab)}$) has the form of $so(2)^{\otimes 3}$ algebra

$$\frac{1}{2}\sum_{n=1}^{3} \alpha_n i\sigma_{n,n}^{(ab)} \to \sum_{n,m=1}^{3} \alpha_n \Lambda_{n,n}^{(ab)} = \begin{pmatrix} 0 & 0 & 0 & -\alpha_1 & 0 & 0 \\ 0 & 0 & 0 & 0 & -\alpha_2 & 0 \\ 0 & 0 & 0 & 0 & 0 & -\alpha_3 \\ \alpha_1 & 0 & 0 & 0 & 0 & 0 \\ 0 & \alpha_2 & 0 & 0 & 0 & 0 \\ 0 & 0 & \alpha_3 & 0 & 0 & 0 \end{pmatrix}. \tag{19}$$

The general SO(6) form of coupling transformation (16) acting on vector $(A_1, A_2, A_3, \tilde{B}_1, \tilde{B}_2, \tilde{B}_3)$ is

$$\mathbf{Y}_{coupling}^{(ab)} = \exp\left(\sum_{n,m=1}^{3} \theta_{nm} \mathbf{\Lambda}_{n,m}^{(ab)}\right) = \begin{pmatrix} \cos\alpha_1 & 0 & 0 & -\sin\alpha_1 & 0 & 0 \\ 0 & \cos\alpha_2 & 0 & 0 & -\sin\alpha_2 & 0 \\ 0 & 0 & \cos\alpha_3 & 0 & 0 & -\sin\alpha_3 \\ \sin\alpha_1 & 0 & 0 & \cos\alpha_1 & 0 & 0 \\ 0 & \sin\alpha_2 & 0 & 0 & \cos\alpha_2 & 0 \\ 0 & 0 & \sin\alpha_3 & 0 & 0 & \cos\alpha_3 \end{pmatrix}. \quad (20)$$

Evolution equation for $\mathbf{Q}^{(3)} = (\mathbf{A}, \tilde{\mathbf{B}})$ in matrix notations is

$$\frac{d}{dt}\begin{pmatrix} \mathbf{A} \\ \tilde{\mathbf{B}} \end{pmatrix} = \begin{pmatrix} 0 & -\gamma \\ \gamma & 0 \end{pmatrix}\begin{pmatrix} \mathbf{A} \\ \tilde{\mathbf{B}} \end{pmatrix}, \quad (21)$$

where matrix $\gamma$ is a set of coefficients in equation (15). Matrix solution of equation (16) for time-independent $\gamma$ is

$$\begin{pmatrix} \mathbf{A}(t) \\ \tilde{\mathbf{B}}(t) \end{pmatrix} = \begin{pmatrix} \cos(t\gamma) & -\sin(t\gamma) \\ \sin(t\gamma) & \cos(t\gamma) \end{pmatrix}\begin{pmatrix} \mathbf{A}_0 \\ \tilde{\mathbf{B}}_0 \end{pmatrix}, \quad (22)$$

Abelian subalgebra $\{i\sigma_{11}^{(ab)}, i\sigma_{22}^{(ab)}, i\sigma_{33}^{(ab)}\}$ plays a special role in classification of entangling gates: in the context of Khaneja-Glaser Decomposition (KGD) it represents a universal set of coupling generators [26-28]. We will be using that feature in the next sections.

A coupling transformation $U = \exp[1/2(i\alpha_1\sigma_{1,1}^{(ab)} + i\alpha_2\sigma_{2,2}^{(ab)} + i\alpha_3\sigma_{3,3}^{(ab)})]$ generated by Abelian Lie algebra (19) corresponds to the following SO(6) matrix

$$\exp\left(\sum_{n=1}^{3} \alpha_n \mathbf{\Lambda}_{n,n}^{(ab)}\right) = \begin{pmatrix} \cos\alpha_1 & 0 & 0 & -\sin\alpha_1 & 0 & 0 \\ 0 & \cos\alpha_2 & 0 & 0 & -\sin\alpha_2 & 0 \\ 0 & 0 & \cos\alpha_3 & 0 & 0 & -\sin\alpha_3 \\ \sin\alpha_1 & 0 & 0 & \cos\alpha_1 & 0 & 0 \\ 0 & \sin\alpha_2 & 0 & 0 & \cos\alpha_2 & 0 \\ 0 & 0 & \sin\alpha_3 & 0 & 0 & \cos\alpha_3 \end{pmatrix} \quad (23)$$

The action of (22) on six-dimensional vector $(A_1, A_2, A_3, \tilde{B}_1, \tilde{B}_2, \tilde{B}_3)$ has a simple geometric form of a set of independent rotations in planes 1-4, 2-5 and 3-6:

$$\begin{pmatrix} A_1 \\ A_2 \\ A_3 \\ \tilde{B}_1 \\ \tilde{B}_2 \\ \tilde{B}_3 \end{pmatrix} \rightarrow \begin{pmatrix} \cos\alpha_1 A_1 - \sin\alpha_1 \tilde{B}_1 \\ \cos\alpha_2 A_2 - \sin\alpha_2 \tilde{B}_2 \\ \cos\alpha_3 A_3 - \sin\alpha_3 \tilde{B}_3 \\ \sin\alpha_1 A_1 + \cos\alpha_1 \tilde{B}_1 \\ \sin\alpha_2 A_2 + \cos\alpha_2 \tilde{B}_2 \\ \sin\alpha_3 A_3 + \cos\alpha_3 \tilde{B}_3 \end{pmatrix} \quad (24)$$

We would like to finish this section by an illustration of how the adjoint action of local transformation modifies so(6) generators $\mathbf{\Lambda}_{n,m}^{(ab)}$. As an example consider $i\sigma_1 \rightarrow i\sigma_2$ rotation of su(2) generator $\sigma_1$ of the first ($a$) qubit by an adjoint $\sigma_3$ action $e^{-i\pi/4\sigma_3^{(a)}} i\sigma_1^{(a)} e^{i\pi/4\sigma_3^{(a)}} = i\sigma_2^{(a)}$. The coupling generator $\sigma_{11}^{(a,b)}$, for example, is transformed as $e^{-i\pi/4\sigma_3^{(a)}} i\sigma_{11}^{(ab)} e^{i\pi/4\sigma_3^{(a)}} = i\sigma_{21}^{(ab)}$. At the group level we have $e^{-i\pi/4\sigma_3^{(a)}} \exp(\xi i\sigma_{1,1}^{(a,b)}) e^{i\pi/4\sigma_3^{(a)}} = \exp(\xi i\sigma_{2,1}^{(a,b)})$. Therefore we expect that the corresponding adjoint transformation of SO(6) operator $\exp(\xi \mathbf{\Lambda}_{1,1}^{(ab)})$ (this is a 1-4 plane rotation), will be identical to 2-4 plane rotation, $\exp(\xi \Lambda_{2,1}^{(ab)})$. Indeed, for the action of $\exp(-\pi/2\, l_{21}^{(a)}) \exp(\xi \Lambda_{1,1}^{(ab)}) \exp(\pi/2\, l_{21}^{(a)})$ on $(A_1, A_2, A_3, \tilde{B}_1, \tilde{B}_2, \tilde{B}_3)$ we have

$$\begin{pmatrix} A_1 \\ A_2 \\ A_3 \\ \tilde{B}_1 \\ \tilde{B}_2 \\ \tilde{B}_3 \end{pmatrix} \to \begin{pmatrix} A_2 \\ -A_1 \\ A_3 \\ \tilde{B}_1 \\ \tilde{B}_2 \\ \tilde{B}_3 \end{pmatrix} \to \begin{pmatrix} A_2 \cos\xi - \tilde{B}_1 \sin\xi \\ -A_1 \\ A_3 \\ \tilde{B}_1 \cos\xi + A_2 \sin\xi \\ \tilde{B}_2 \\ \tilde{B}_3 \end{pmatrix} \to \begin{pmatrix} A_1 \\ A_2 \cos\xi - \tilde{B}_1 \sin\xi \\ A_3 \\ \tilde{B}_1 \cos\xi + A_2 \sin\xi \\ \tilde{B}_2 \\ \tilde{B}_3 \end{pmatrix} \equiv e^{\xi \Lambda_{2,1}^{(ab)}} \begin{pmatrix} A_1 \\ A_2 \\ A_3 \\ \tilde{B}_1 \\ \tilde{B}_2 \\ \tilde{B}_3 \end{pmatrix}. \tag{25}$$

In the following sections we will be relying on the concept that coupling generators $\sigma_{nm}^{(ab)}$ can be modified by the adjoint action of local transformations $\sigma_{nm}^{(ab)} \to \sigma_{n'm'}^{(ab)}$ and consequently local transformations can be used to change rotation $\Lambda_{n,m}^{(ab)}$ to $\Lambda_{n',m'}^{(ab)}$.

## VII. SO(6) FORM OF SU(4) INVARIANTS FOR QUBIT-QUBIT COUPLING

Qubit-qubit coupling operations belong to SU(4) group. A trivial invariant of SU(4) group is Hermitian inner product which is simply a norm of the wave function. On the other hand, $SU(4)^{(a,b)}$ coupling between qubits $a$ and $b$, for example, does not affect density matrix $\rho^{(c)} = Tr_{(ab)} |\psi^{(abc)}\rangle\langle\psi^{(abc)}| = 1/2(\mathbf{I}^{(c)} + \vec{m}^{(c)}\vec{\sigma})$. Local $SU(2)^{(c)}$ transformations on qubit $c$ result in rotations of Bloch vector $\vec{m}^{(c)}$. Since $\vec{m}^{(c)} \cdot \vec{m}^{(c)}$ is invariant under this action we have $Det(\rho^{(c)}) = 1/4(1 - \vec{m}^{(c)} \cdot \vec{m}^{(c)})$ as invariant of both $SU(4)^{(a,b)}$ and $SU(2)^{(c)}$ actions. Determinant is related to concurrence $C^{(c)}$, as defined in equation (6) of ref. [29], $Det(\rho^{(c)}) = (1/2\, C^{(c)})^2$. In this section we demonstrate that this invariant naturally emerges as an invariant of $SO(6)^{(a,b)}$ coupling transformations of vectors $\mathbf{B}$ and $\mathbf{C}$.

Canonical bilinear invariant form of SO(6) action is a 6-dimensional dot product $(\mathbf{A}, \tilde{\mathbf{B}}) \cdot (\mathbf{A}, \tilde{\mathbf{B}}) = \mathbf{A}^2 - \mathbf{B}^2$, however due to Plücker relations (9) this invariant takes a trivial form $\mathbf{A}^2 - \mathbf{B}^2 \equiv 0$. Another SO(6) invariant is a Hermitian inner product: since SO(6) is a subgroup of $SU(6)$ we have real-valued invariant $(\mathbf{A}, \tilde{\mathbf{B}})(\mathbf{A}, \tilde{\mathbf{B}})^* = \mathbf{A}\mathbf{A}^* + \mathbf{B}\mathbf{B}^*$. This invariant does not change under qubits $a$, $b$ coupling and local transformations on qubit $c$. Therefore, this invariant must be associated with bi-partite concurrence in the system $c(ab)$. Indeed, we have the following relation for (squared) concurrences which are defined as $\tau_{a(bc),b(ca),c(ab)} = 4 \det \rho^{(a,b,c)}$

$$\tau_{c(ab)} = 2(\mathbf{A}^* \cdot \mathbf{A} + \mathbf{B}^* \cdot \mathbf{B}), \; \tau_{a(bc)} = 2(\mathbf{B}^* \cdot \mathbf{B} + \mathbf{C}^* \cdot \mathbf{C}), \; \tau_{b(ca)} = 2(\mathbf{C}^* \cdot \mathbf{C} + \mathbf{A}^* \cdot \mathbf{A}), \tag{26}$$

Notice that Coffman-Kundu-Wootters (equation (24) in ref. [24]) immediately follows from eq (10,11,24),

$$\tau_{(bc)} + \tau_{(ac)} = 2(\mathbf{A} \cdot \mathbf{A}^* + \mathbf{B} \cdot \mathbf{B}^* - |\mathbf{A} \cdot \mathbf{A}| - |\mathbf{B} \cdot \mathbf{B}|) \equiv \tau_{c(ab)} - \tau_{abc}, \tag{27}$$

As we discussed above $\tau_{c(ab)}$ is an invariant of local and coupling transformations between $a$ and $b$ qubits such that $\tau_{c(ab)}$ does not change if qubits $a$ and $b$ are being coupled and equation (27) suggests that entanglement satisfies some form of conservation rule. Let's rewrite this equation as $\tau_{c(ab)} = \tau_{abc} + \tau_{(bc)} + \tau_{(ac)}$. If transformations are limited to qubits $a$ and $b$ including coupling operations, i.e. full $SU(4)^{(a,b)}$ group, the sum of three-tangle and two-tangles does not change $\tau_{abc} + \tau_{(bc)} + \tau_{(ac)} = const$ because $\tau_{c(ab)}$ does not change. The question is whether $\tau_{(bc)}$ and $\tau_{(ac)}$ can be used as a resource for boosting, or minimizing, three-tangle, or, alternatively, whether $\tau_{abc}$ and $\tau_{(bc)}$ can be transformed into entanglement between $a$ and $c$ qubits. For example, if the bipartite entanglements $\tau_{(bc)}$ between the $b$ and $c$ qubits (equation (11a)), and $\tau_{(ac)}$ between $a$ and $c$ qubits (equation (11b)) can be made zero by manipulating vectors $\mathbf{A}$ and $\mathbf{B}$, defining $\tau_{(bc)}$ and $\tau_{(ac)}$ via eq. (11b,c), then the maximum achievable value of the three tangle by $SU(4)^{(a,b)}$ operations is given by the value of the bipartite entanglement between qubit $c$ and the composite subsystem $(ab)$, i.e. $\tau_{abc}^{\max} = \tau_{c(ab)}$. Similarly, for $\tau_{abc} = 0$ and $\tau_{(bc)} = 0$ we have $\tau_{(ac)}^{\max} = \tau_{c(ab)}$. As we will show in the following sections these operations are indeed accessible by $SU(4)^{(a,b)}$ transformations and $SO(6)^{(a,b)}$ geometry can be efficiently used to design required transformations in an optimal way.

## VIII. EXAMPLES OF SU(4) GATES: CNOT, CZ, SWAP

To illustrate the relation between quantum transformations induced by two-qubit coupling in the $SU(4)$ and $SO(6)$ forms we analyze two classes of entangling gates: CZ-type and SWAP-type gates.

Using KGD decomposition [27,28] entangling gates can be partitioned into local equivalence classes: gates which are different only by local transformations belong to the same entangling class. $CZ^{(ab)} = |0^{(a)}\rangle\langle 0^{(a)}| \otimes \mathbf{I}^{(b)} + |1^{(a)}\rangle\langle 1^{(a)}| \otimes \sigma_z^{(b)}$ and $CNOT^{(ab)} = |0^{(a)}\rangle\langle 0^{(a)}| \otimes \mathbf{I}^{(b)} + |1^{(a)}\rangle\langle 1^{(a)}| \otimes \sigma_x^{(b)}$ belong to the same local equivalence class. These gates also belong to the controlled-U class of coupling gates.

It is important to realize that relation between $SU(4)$ and $SO(6)$ transformations algebraically are implemented by a map between corresponding generators. If the $SU(4)$ transformation is given in the form of exponents of $su(4)$ generators then $SO(6)$ transformation is given by exponentiation of corresponding combination of $so(6)$ generators according to equations (7) and (17). CZ and CNOT gates are generated by combinations of global phase $\exp(i\pi/2)$, local rotations and coupling terms $\sigma_{zz}^{(ab)}$ and $\sigma_{zx}^{(ab)}$

$$CZ = e^{i\pi/4 + i\pi/4(\sigma_{zz}^{(ab)} - \sigma_z^{(b)} - \sigma_z^{(a)})} \equiv e^{i\pi/4} e^{i\pi/4 \sigma_{zz}^{(ab)}} e^{-i\pi/4 \sigma_z^{(b)}} e^{-i\pi/4 \sigma_z^{(a)}}$$
$$CNOT = e^{i\pi/4 + i\pi/4(\sigma_{zx}^{(ab)} - \sigma_z^{(b)} - \sigma_x^{(a)})} \equiv e^{i\pi/4} e^{i\pi/4 \sigma_{zx}^{(ab)}} e^{-i\pi/4 \sigma_z^{(b)}} e^{-i\pi/4 \sigma_x^{(a)}}$$
(28)

Both gates belong to the same KGD class. We have already discussed how local rotations act on $\mathbf{A}$, $\mathbf{B}$, $\mathbf{C}$ vectors and here we demonstrate the action of coupling operators.

Local operators $\mathbf{U}_{local}^{(a)}$ and $\mathbf{U}_{local}^{(b)}$ are acting separately (independently) on vectors $\mathbf{A}$ and $\mathbf{B}$ as $SO(3)$ matrices (see section III above). For example, $\exp(-i\pi/4\sigma_z^{(a)}) \Box \exp(-\pi/2 l_{21}^{(a)})$ has the following effect on $\mathbf{A}$:

$$(A_1, A_2, A_3) \to (-A_2, A_1, A_3).$$
(29)

Operator $\exp(-i\pi/4\sigma_x^{(b)}) \Box \exp(-\pi/2 l_{32}^{(b)})$ rotates vector $\tilde{\mathbf{B}}$:

$$(\tilde{B}_1, \tilde{B}_2, \tilde{B}_3) \to (\tilde{B}_1, -\tilde{B}_3, \tilde{B}_2).$$
(30)

Coupling operator $\exp(i\pi/4\sigma_{zz}^{(ab)})$ acts as $\exp(\pi/2\Lambda_{3,3}^{(ab)})$. According to equation (19) this operator is a $\pi/2$ rotation involving $\tilde{B}_3$ and $A_3$ components of $\mathbf{A}$ and $\tilde{\mathbf{B}}$ while other components remain unchanged, $A_3 \to -\tilde{B}_3$, $\tilde{B}_3 \to A_3$. Action of $CZ^{(ab)}$ gate is a combination of two local rotations (29, 30) and coupling rotation $\exp(\pi/2\Lambda_{3,3}^{(ab)})$ and multiplication by phase factor $\exp(i\pi/2)$

$$\begin{pmatrix} A_1 \\ A_2 \\ A_3 \\ \tilde{B}_1 \\ \tilde{B}_2 \\ \tilde{B}_3 \end{pmatrix} \to \begin{pmatrix} -A_2 \\ A_1 \\ A_3 \\ -\tilde{B}_2 \\ \tilde{B}_1 \\ \tilde{B}_3 \end{pmatrix} \to \begin{pmatrix} -A_2 \\ A_1 \\ -\tilde{B}_3 \\ -\tilde{B}_2 \\ \tilde{B}_1 \\ A_3 \end{pmatrix} \xrightarrow{(\times i)} \begin{pmatrix} -iA_2 \\ iA_1 \\ -i\tilde{B}_3 \\ -i\tilde{B}_2 \\ i\tilde{B}_1 \\ iA_3 \end{pmatrix} \quad (31)$$

The action $CNOT^{(ab)}$ is nearly identical to $CZ^{(ab)}$ with the coupling action $\exp(\pi/2 \Lambda_{3,1}^{(ab)})$

$$\begin{pmatrix} A_1 \\ A_2 \\ A_3 \\ \tilde{B}_1 \\ \tilde{B}_2 \\ \tilde{B}_3 \end{pmatrix} \to \begin{pmatrix} -A_2 \\ A_1 \\ A_3 \\ \tilde{B}_1 \\ -\tilde{B}_3 \\ \tilde{B}_2 \end{pmatrix} \to \begin{pmatrix} -A_2 \\ A_1 \\ -\tilde{B}_1 \\ A_3 \\ \tilde{B}_3 \\ \tilde{B}_2 \end{pmatrix} \xrightarrow{(\times i)} \begin{pmatrix} -iA_2 \\ iA_1 \\ -i\tilde{B}_1 \\ iA_3 \\ i\tilde{B}_3 \\ i\tilde{B}_2 \end{pmatrix} \quad (32)$$

Remarkably, in both cases local rotations do not involve those components of vectors $\mathbf{A}$ and $\tilde{\mathbf{B}}$ which are being affected by coupling transformations because local generators involved in these gates commute with the coupling generators. Also, notice that action (31) is identical to transformation (32) if one performs pre- and post- rotations $\exp(\pi/2 \, I_{31}^{(b)})$ which swaps components $\tilde{B}_1$ and $\tilde{B}_3$.

Another ubiquitous coupling operator is the SWAP gate. Its action on two qubit state has the following form:

$$SWAP_{su(4)} = \begin{pmatrix} 1 & 0 & 0 & 0 \\ 0 & 0 & 1 & 0 \\ 0 & 1 & 0 & 0 \\ 0 & 0 & 0 & 1 \end{pmatrix} = \exp\left\{ i\frac{\pi}{4}(\sigma_{xx} + \sigma_{yy} + \sigma_{zz}) - i\frac{\pi}{4} \right\} \quad (33)$$

Corresponding action on vectors $\mathbf{A}$ and $\tilde{\mathbf{B}}$ is $\exp(\pi/2 \Lambda_{11}^{(ab)}) \exp(\pi/2 \Lambda_{22}^{(ab)}) \exp(\pi/2 \Lambda_{33}^{(ab)})$, which is the $\mathbf{A} \leftrightarrow (\pm 1)\tilde{\mathbf{B}}$ swap and multiplication by $\exp(i\pi/2)$:

$$\begin{pmatrix} \mathbf{A} \\ \tilde{\mathbf{B}} \end{pmatrix} \to \begin{pmatrix} \tilde{\mathbf{B}} \\ -\mathbf{A} \end{pmatrix} \xrightarrow{(\times i)} \begin{pmatrix} i\tilde{\mathbf{B}} \\ -i\mathbf{A} \end{pmatrix} \quad (34)$$

The minus sign in this equation is a signature of an SO-type matrix having determinant =1.

### IX. GEOMETRIC CONSTRUCTION OF AN ARBITRARY TWO-QUBIT TRANSFORMATION BY CONCATENATING THREE CZ GATES

As a simple application of the geometric technique, we suggest the following practical quantum control problem. One wants to generate an arbitrary two-qubit coupling gate by essentially applying three fixed coupling-strength CZ-type gates. The coupling term in a general KGD decomposition has the form of a three-parameter transformation:

$$Two\,Qubit\,Coupling = \exp\left( 1/2 \sum_{n=1}^{3} \alpha_n i \sigma_{n,n}^{(ab)} \right) \quad (35)$$

The geometric SO(6) form of this transformation is given by equation (19).

Suggested solution of this problem [30-33] does not offer a clear insight on the method of how this solution was obtained, which make generalization of this result somewhat unobvious. A geometric solution described below allows to design the complete set of required transformations in the form of elementary two-dimensional rotations of six-dimensional vector $(\mathbf{A},\tilde{\mathbf{B}})$. Designing a set of local transformations in combination with coupling operators becomes a matter of trivial geometric manipulations of vector $(\mathbf{A},\tilde{\mathbf{B}})$. It will also become clear that optimization of coupling transformations is determined by the proper choice of auxiliary local transformations. While it goes beyond the scope of this paper, one can apply geometric method for intuitive analysis of robustness of the whole transformation relative to possible errors at intermediate steps.

Since our goal is to demonstrate that geometry provides useful insight into three-qubit control problem we will be using a set of coupling gates locally equivalent to CZ gate instead of the exact CZ gate, which would have obscured the main idea. Coupling operators from the CZ-class have the form of transformations $\exp(\pi/4 i\sigma_{nm})$, where parameter defining the coupling strength is fixed to be to $\pi/4$ (in general, coupling operators $\exp(i\varsigma\sigma_{\tilde{n},\tilde{m}})$ are locally equivalent to controlled-U gates [33]).

At first glance, obtaining three free-parameter transformation (35) seems to be impossible. However, this task has a simple solution strategy if one applies SO(6) geometry. Notice that any CZ-class transformation is simply swapping $\mathbf{A}_n$ and $\tilde{\mathbf{B}}_m$ components of $(\mathbf{A},\tilde{\mathbf{B}})$ vector. Therefore such gates can be used to move one of $\mathbf{A}_n$ components to the 4-6 position of $(\mathbf{A},\tilde{\mathbf{B}})$ while $\tilde{\mathbf{B}}_m$ is moved to the first three positions of this vectors. Then one of two $\mathbf{A}_k$ components can be mixed with $\tilde{\mathbf{B}}_m$ by local rotation containing coupling parameter $\alpha_1$ (as in equations (14)). Next, the proper linear combinations of $\mathbf{A}_k$ and $\tilde{\mathbf{B}}_m$ are swapped back to CZ-class gate to match equation (23).

According to this strategy we perform first the local $A_2 \to \tilde{B}_1$ component swap by

$$\exp(\pi/4 i\sigma_{21}^{(ab)}) \square \exp(\pi/2 \Lambda_{21}^{(ab)}). \tag{36}$$

Then, local transformation on qubit $a$ mixes $A_1$ and $\tilde{B}_1$ by

$$\exp(-\alpha_1/2 i\sigma_z^{(a)}) \square \exp(-\alpha_1 l_{21}^{(a)}). \tag{37}$$

Similarly mixing $A_2$ and $\tilde{B}_2$ can be achieved by local transformations on qubit $b$

$$\exp(\alpha_2/2 i\sigma_z^{(b)}) \square \exp(\alpha_2 l_{21}^{(b)}), \tag{38}$$

such that we have the following geometric sequence

$$\begin{pmatrix} A_1 \\ A_2 \\ A_3 \\ \tilde{B}_1 \\ \tilde{B}_2 \\ \tilde{B}_3 \end{pmatrix} \to \begin{pmatrix} A_1 \\ -\tilde{B}_1 \\ A_3 \\ A_2 \\ \tilde{B}_2 \\ \tilde{B}_3 \end{pmatrix} \to \begin{pmatrix} A_1 \cos\alpha_1 - \tilde{B}_1 \sin\alpha_1 \\ -\tilde{B}_1 \cos\alpha_1 - A_1 \sin\alpha_1 \\ A_3 \\ A_2 \cos\alpha_2 - \tilde{B}_2 \sin\alpha_1 \\ \tilde{B}_2 \cos\alpha_2 + A_2 \sin\alpha_2 \\ \tilde{B}_3 \end{pmatrix} \tag{38}$$

Next we mix $A_3$ and $\tilde{B}_3$. For that purpose $A_3$ and $\tilde{B}_3$ have to appear either in 1-3 or 4-6 sets of component of $(\mathbf{A},\tilde{\mathbf{B}})$ vector. Let's say we choose the second route. A CZ-type coupling

$$\exp(\pi/4 i\sigma_{31}^{(ab)}) \square \exp(\pi/2 \Lambda_{31}^{(ab)}) \tag{39}$$

is concatenated with local rotation

$$\exp(-\alpha_3/2 i\sigma_2^{(b)}) \square \exp(\alpha_3 l_{31}^{(b)}) \tag{40}$$

on qubit $b$

$$\begin{pmatrix} A_1 \cos\alpha_1 - \tilde{B}_1 \sin\alpha_1 \\ -\tilde{B}_1 \cos\alpha_1 - A_1 \sin\alpha_1 \\ A_3 \\ A_2 \cos\alpha_2 - \tilde{B}_2 \sin\alpha_1 \\ \tilde{B}_2 \cos\alpha_2 + A_2 \sin\alpha_2 \\ \tilde{B}_3 \end{pmatrix} \to \begin{pmatrix} A_1 \cos\alpha_1 - \tilde{B}_1 \sin\alpha_1 \\ -\tilde{B}_1 \cos\alpha_1 - A_1 \sin\alpha_1 \\ -A_2 \cos\alpha_2 + \tilde{B}_2 \sin\alpha_1 \\ A_3 \\ \tilde{B}_2 \cos\alpha_2 + A_2 \sin\alpha_2 \\ \tilde{B}_3 \end{pmatrix} \to \begin{pmatrix} A_1 \cos\alpha_1 - \tilde{B}_1 \sin\alpha_1 \\ -\tilde{B}_1 \cos\alpha_1 - A_1 \sin\alpha_1 \\ -A_2 \cos\alpha_2 + \tilde{B}_2 \sin\alpha_1 \\ A_3 \cos\alpha_3 - \tilde{B}_1 \sin\alpha_3 \\ \tilde{B}_2 \cos\alpha_2 + A_2 \sin\alpha_2 \\ \tilde{B}_3 \cos\alpha_3 + A_3 \sin\alpha_3 \end{pmatrix} \qquad (41)$$

Third coupling transformation swaps $(-\tilde{B}_1 \cos\alpha_1 - A_1 \sin\alpha)$ and $(A_3 \cos\alpha_3 - \tilde{B}_1 \sin\alpha_3)$ by CZ-type operator

$$\exp\left(\pi/4 i\sigma_{21}^{(ab)}\right) \square \exp\left(\pi/2 \Lambda_{21}^{(ab)}\right) \qquad (42)$$

which is followed by a trivial local rotation

$$\exp\left(\pi/4\, i\sigma_1^{(a)}\right) \square \exp\left(\pi/2\, l_{32}^{(a)}\right) \qquad (43)$$

The transformation is complete:

$$\begin{pmatrix} A_1 \cos\alpha_1 - \tilde{B}_1 \sin\alpha_1 \\ -\tilde{B}_1 \cos\alpha_1 - A_1 \sin\alpha_1 \\ -A_2 \cos\alpha_2 + \tilde{B}_2 \sin\alpha_1 \\ A_3 \cos\alpha_3 - \tilde{B}_1 \sin\alpha_3 \\ \tilde{B}_2 \cos\alpha_2 + A_2 \sin\alpha_2 \\ \tilde{B}_3 \cos\alpha_3 + A_3 \sin\alpha_3 \end{pmatrix} \to \begin{pmatrix} A_1 \cos\alpha_1 - \tilde{B}_1 \sin\alpha_1 \\ A_3 \cos\alpha_3 - \tilde{B}_1 \sin\alpha_3 \\ -A_2 \cos\alpha_2 + \tilde{B}_2 \sin\alpha_1 \\ \tilde{B}_1 \cos\alpha_1 + A_1 \sin\alpha_1 \\ \tilde{B}_2 \cos\alpha_2 + A_2 \sin\alpha_2 \\ \tilde{B}_3 \cos\alpha_3 + A_3 \sin\alpha_3 \end{pmatrix} \to \begin{pmatrix} A_1 \cos\alpha_1 - \tilde{B}_1 \sin\alpha_1 \\ A_2 \cos\alpha_2 - \tilde{B}_2 \sin\alpha_2 \\ A_3 \cos\alpha_3 - \tilde{B}_1 \sin\alpha_3 \\ \tilde{B}_1 \cos\alpha_1 + A_1 \sin\alpha_1 \\ \tilde{B}_2 \cos\alpha_2 + A_2 \sin\alpha_2 \\ \tilde{B}_3 \cos\alpha_3 + A_3 \sin\alpha_3 \end{pmatrix} \qquad (44a)$$

Transformation (44a) is identical to transformation (23) which is the SO(6) representation of SU(4) operator (35).

To summarize, transformation $\exp\left(1/2 \sum_{n=1}^{3} \alpha_n i\sigma_{n,n}^{(ab)}\right)$ is constructed as a combination of operators (36)-(43)

$$\exp\left(\frac{\pi}{4}i\sigma_1^{(a)}\right)\exp\left(\frac{\pi}{4}i\boldsymbol{\sigma}_{21}^{(ab)}\right)\exp\left(-\frac{\alpha_3}{2}i\boldsymbol{\sigma}_2^{(b)}\right)\exp\left(\frac{\pi}{4}i\boldsymbol{\sigma}_{31}^{(ab)}\right)\exp\left(\frac{\alpha_2}{2}i\sigma_3^{(b)}\right)\exp\left(-\frac{\alpha_2}{2}i\sigma_3^{(a)}\right)\exp\left(\frac{\pi}{4}i\boldsymbol{\sigma}_{21}^{(ab)}\right) \qquad (44b)$$

Three coupling terms are highlighted in bold font.

## X. TRANSFORMATION BETWEEN ASYMMETRIC W AND GHZ STATES

It is well known that W and GHZ states are locally inequivalent, i.e. there is no sequence of local SU(2) transformations which transforms **W** state into **GHZ** (and vice versa). The transformation we want to design necessarily involves qubit-qubit coupling. It is also true that such coupling cannot be implemented by a single $a$-$b$, $b$-$c$ or $a$-$c$ interaction: it necessarily involves a sequence of two couplings applied between two pairs of qubits. An example of such transformation for a standard W state was suggested in [1]. It involved a sequence of two $b$-$c$ couplings and one $a$-$c$ coupling. Here we generalize this result to a wider class of states, which we call asymmetric **W** states. We also optimized the coupling transformations by eliminating one $b$-$c$ coupling transformation.

We define a class of normalized asymmetric **W** states as $|\mathbf{W}\rangle = \alpha|0,0,1\rangle + \beta|0,1,0\rangle + \gamma|1,0,0\rangle$. By local rotations and global phase manipulation coefficients $\alpha$, $\beta$ and $\gamma$ can be made purely real. To guarantee that the state is normalized we use spherical-angle representation similar to equation (20) in [34].

$$|\mathbf{W}\rangle = \sin\theta\cos\varphi|0,0,1\rangle + \sin\theta\sin\varphi|0,1,0\rangle + \cos\theta|1,0,0\rangle. \qquad (45)$$

Vectors **A**, **B** and **C** evaluate to

$$\mathbf{A}_W = \sin\theta^2 \sin\varphi\cos\varphi \times \boldsymbol{\mu}, \quad \mathbf{B}_W = \sin\theta\cos\theta\cos\varphi \times \boldsymbol{\mu}, \quad \mathbf{C}_W = \sin\theta\cos\theta\sin\varphi \times \boldsymbol{\mu}, \tag{46}$$

where $\boldsymbol{\mu} = (i,-1,0)$. Notice that if one sets $\theta = 0$ state (45) becomes a completely separable state and all three vectors defined by eq. (46) evaluate to zero. One obtains bi-separable state by taking $\phi = 0$, $|W\rangle = (\sin\theta|0,1\rangle_{ac} + \cos\theta|1,0\rangle_{ac}) \otimes |0\rangle_b$. Standard symmetric W state is recovered by taking $\phi = \pi/4$ and $\theta = \arctan\sqrt{2} \equiv \arccos(1/\sqrt{3})$. Vector $\boldsymbol{\mu}$ equation (46) satisfies $\boldsymbol{\mu}^2 = 0$, $\boldsymbol{\mu}\boldsymbol{\mu}^* = 2$, and according to equation (10), (11a,b,c) three-tangle for $|W\rangle$ state (45) is zero $\tau_{abc}^W = 0$, and two-tangles are $\tau_{(ab)}^W = 4(\sin\theta\cos\theta\sin\varphi)^2$, $\tau_{(ac)}^W = 4(\sin\theta\cos\theta\cos\varphi)^2$, $\tau_{(bc)}^W = 4(\sin\theta^2\sin\varphi\cos\varphi)^2$. One interesting physical property of the asymmetric state $|W\rangle$ is that $\tau_{a(bc)}^W = (\sin 2\theta)^2$. Maximal bi-partite concurrence therefore is $\text{Max}(\tau_{a(bc)}^W) = 1$ for $\theta = \pi/4$. The state (45) is $|W\rangle = (\cos\varphi|0,0,1\rangle + \sin\varphi|0,1,0\rangle + |1,0,0\rangle)/\sqrt{2}$. For this state $\tau_{(ab)}^W = \sin\varphi^2$, $\tau_{(ac)}^W = \cos\varphi^2$ such that CKW equation (27) $\tau_{abc} = \tau_{a(bc)} - \tau_{ab} - \tau_{ac}$ has a trivial form $0 = 1 - \sin\phi^2 - \cos\phi^2$. As we will explain in the next section the entanglement resource of $\tau_{ab}$ and $\tau_{ac}$ can be converted by b-c coupling into three-tangle resource and vice versa.

To simplify equations for GHZ vectors we introduce an additional global phase

$$|\text{GHZ}\rangle = 1/\sqrt{2}\, e^{-i\pi/4}(|0,0,0\rangle + |1,1,1\rangle). \tag{47}$$

Vectors

$$\mathbf{A}_{\text{GHZ}}, \mathbf{B}_{\text{GHZ}}, \mathbf{C}_{\text{GHZ}} = (0,0,1)/2 \tag{48}.$$

Three-tangle, and two-tangles for $|\text{GHZ}\rangle$ state are $\tau_{abc}^{\text{GHZ}} = 1$, $\tau_{(ab),(bc),(ac)}^{\text{GHZ}} = 0$, $\tau_{c(ab),a(bc),b(ac)}^{\text{GHZ}} = 1$.

First we design $b-c$ coupling operation by comparing vectors $(\mathbf{B},\tilde{\mathbf{C}})_W$, and $(\mathbf{B},\tilde{\mathbf{C}})_{\text{GHZ}}$ (here $\tilde{\mathbf{C}} = -i\mathbf{C}$)

$$(\mathbf{B},\tilde{\mathbf{C}})_W = \sin\theta\cos\theta \times (i\cos\varphi, -\cos\varphi, 0, \sin\varphi, i\sin\varphi, 0)$$
$$(\mathbf{B},\tilde{\mathbf{C}})_{\text{GHZ}} = (0,0,1/2,0,0,-i/2) \tag{49a,b}$$

General rules of SO(6) control on vector $(\mathbf{B},\tilde{\mathbf{C}})$ are as follows: 1) real and imaginary components will transform into real and imaginary components correspondingly because SO(6) are real-valued matrices, 2) local operations on $b$ and $c$ are cost-free and these operations act as a complete sets of SO(3) rotations on 1-3 and 4-6 components correspondingly, 3) qubit-qubit coupling mixes components 1-3 with components 4-6.

To eliminate, simultaneously, imaginary component $i\cos\varphi$ of $\mathbf{B}_W$ (compare equations (49(ab))) and real-valued component $\sin\varphi$ of $\tilde{\mathbf{C}}_W$, i.e. imaginary component of $\mathbf{C}_W = i\tilde{\mathbf{C}}_W$, we implement a $\pi/2$ rotation in the 1-4 plane of $(\mathbf{B},\tilde{\mathbf{C}})_W$. As we noted above this operation is a CZ-class gate, implemented by the coupling operator $\exp(i\pi/4\sigma_{xx}^{bc}) \Box \exp(\pi/2\Lambda_{11})$. The transformed state

$$|W_1\rangle = \exp(i\pi/4\sigma_{xx})|W\rangle = \frac{1}{\sqrt{2}}\left(e^{i\phi}\sin\theta|001\rangle + ie^{-i\phi}\sin\theta|010\rangle + \cos\theta|100\rangle + i\cos\theta|111\rangle\right) \tag{50}$$

For the state $|W_1\rangle$ we have

$$(\mathbf{B},\tilde{\mathbf{C}})_{W_1} = \sin\theta\cos\theta \times (-\sin\varphi, -\cos\varphi, 0, i\cos\varphi, i\sin\varphi, 0) \tag{51a}$$

Notice that while operation $\exp(i\pi/4\,\sigma_{11}^{bc})$ does not affect $\tau_{a(bc)}$ it boosts three-tangle $\tau_{abc}$ by using of $\tau_{(ac)}$ and $\tau_{(ab)}$ as a resource. New $\tau_{(ac)}$ and $\tau_{(ab)}$ evaluate to zero while $\tau_{abc} = (\sin 2\theta)^2$ (see also section dedicated to three-tangle control below).

To assess the efficiency of this transformation it is helpful to check how much does the distance between states, estimated as Fubini-Study angle

$$\Theta^{FS}(\psi_1, \psi_2) = \underset{SU(2)^{(a,b,c)}}{\text{Min}} \left[ \cos^{-1}\left( \left| \langle \psi_1 | SU(2)^{(a)} \otimes SU(2)^{(b)} \otimes SU(2)^{(c)} | \psi_2 \rangle \right| \right) \right],$$

change when we perform first transformation. For standard $\mathbf{W}$ state $1/\sqrt{3}(|0,0,1\rangle + |0,1,0\rangle + |1,0,0\rangle)$, we have angle $\Theta^{FS}(\mathbf{W}, \mathbf{GHZ})$ between $\mathbf{GHZ}$ and $\mathbf{W}$ decreasing from $45°$ to $\Theta^{FS}(\mathbf{W}_1, \mathbf{GHZ}) \approx 9.7356°$, where $|\mathbf{W}_1\rangle = \frac{1}{\sqrt{6}}\left(e^{i\pi/4}\sqrt{2}|001\rangle + i e^{-i\pi/4}\sqrt{2}|010\rangle + |100\rangle + i|111\rangle\right)$, equation (50). For bi-separable state $\theta = \pi/4$, $\phi = 0$, $|\mathbf{W}^{bs}\rangle = 1/\sqrt{2}(|0,1\rangle_{ac} + |1,0\rangle_{ac}) \otimes |0\rangle_b$ and we have $|\mathbf{W}_1^{bs}\rangle = \frac{1}{2}(|001\rangle + i|010\rangle + |100\rangle + i|111\rangle)$. Angle $\Theta^{FS}$ is decreasing from $\Theta^{FS}(\mathbf{W}^{bs}, \mathbf{GHZ}) = 45°$ to $\Theta^{FS}(\mathbf{W}_1^{bs}, \mathbf{GHZ}) = 0°$, which indicates that after application of the first coupling operator the state is locally equivalent to $\mathbf{GHZ}$. Another example is the state $|\mathbf{W}^{rnd}\rangle = \alpha|0,0,1\rangle + \beta|0,1,0\rangle + \gamma|1,0,0\rangle$ with random coefficients $\alpha = 0.2175$, $\beta = 0.7778$, $\gamma = 0.5895$. We have $\Theta^{FS}$-angle changing as $\Theta^{FS}(\mathbf{W}^{rnd}, \mathbf{GHZ}) = 37.58° \rightarrow \Theta^{FS}(\mathbf{W}_1^{rnd}, \mathbf{GHZ}) = 8.87°$, where $|\mathbf{W}_1^{rnd}\rangle \approx (0.5413 + i0.1821)|001\rangle + (-0.1821 + i0.5413)|010\rangle + 0.5895|100\rangle + i0.5895|111\rangle$.

After first transformation we have the state $|\mathbf{W}_1\rangle$ given by equation (50). Expressions for vectors $\mathbf{B}_{W_1}$ and $\mathbf{C}_{W_1} \equiv i\tilde{\mathbf{C}}_{W_1}$, as defined by equation (51) apparently match the expected result of application of SO(6) rotation $\exp(\pi/2\,\Lambda_{11})$.

New vector $\mathbf{A}$ calculated using definition (3a) for the state $|\mathbf{W}_1\rangle$ is $\mathbf{A}_{W_1} = 1/2(-1, i\cos 2\theta, 0)$. Apparently initial $\mathbf{A}_W$ in equation (46) differs from $\mathbf{A}_{W_1}$. Notice that while $\mathbf{A}$ does not change under any local SL(2) transformation on qubits $b$ and $c$, this vector will change under b-c coupling. This is quite evident from the fact that vector $\mathbf{A}$ defines concurrence $\tau_{(bc)}$ in equation (11a), as well as three-tangle $\tau_{abc}$ in equation (10). These parameters change under b-c coupling, i.e. b-c coupling affects vector $\mathbf{A}$. The set of 9 polynomials $\mathbf{A}$, $\mathbf{B}$, $\mathbf{C}$, defined in (3a,b,c), is a special subset of $8 \times (8+1)/2 = 36$ quadratic polynomials in eight coefficients $c_{i,j,k}$. While the space spanned by $\mathbf{B}$ and $\mathbf{C}$ is a 6-dimensional invariant space relative to $SU(4)^{(bc)}$ group of transformations, the same $SU(4)^{(bc)}$ operators will couple vector $\mathbf{A}$ with other polynomials in the remaining set of $36 - 9 = 27$ polynomials. Mathematically, the coupling has a structure of vector-tensor coupling.

Next obvious local trim operations, which are logical to apply at this stage, are rotation by angle $-\varphi$ in the 1-2 plane followed by $\varphi$-rotation in the 4-5 planes. Corresponding $SU(2)^{b,c}$ operators are $\exp(-i\varphi/2\,\sigma_z^b) \square \exp(-\varphi l_{21}^b)$ and $\exp(i\varphi/2\,\sigma_z^c) \square \exp(\varphi l_{21}^c)$. The state

$$|\mathbf{W}_2\rangle = e^{i\varphi/2\sigma_3^c} e^{-i\varphi/2\sigma_3^b} |W_1\rangle = 1/\sqrt{2}\left(\sin\theta\langle 001| + i\sin\theta\langle 010| + \cos\theta\langle 100| + i\cos\theta\langle 111|\right) \tag{53}$$

Interestingly, this state does not depend any more on parameter $\varphi$. It in not evident from expression (50) that $\varphi$ can be eliminated by a sequence of two local transformation, and it is not obvious which local transformations will eliminate that parameter. But, equation (51) for vectors $\mathbf{B}$ and $\tilde{\mathbf{C}}$ offers an immediate solution for this problem due to better visualization of the three-qubit state in the form of a six-dimensional vector in comparison to a set of 8 complex coefficients $c_{i,j,k}$.

As expected, new vectors

$$\left(\mathbf{B}, \tilde{\mathbf{C}}\right)_{W_2} = \frac{\sin 2\vartheta}{2\sqrt{2}}(0, -1, 0, i, 0, 0) . \tag{54}$$

As it clearly follows from comparison of (49b) and (54), further application of b-c coupling will not make the state (49a) any closer to **GHZ** state. Now we apply coupling in the *a*-*b* (or *a*-*c*) pair. The *a*-*b* coupling is acting as an SO(6) transformation in partition 3, i.e. it affects vector $\left(\mathbf{A}, \tilde{\mathbf{B}}\right)$. After first coupling operation resulting in equation (50) and local rotation (53) we have

$$\left(\mathbf{A}, \tilde{\mathbf{B}}\right)_{W_2} = \frac{1}{2\sqrt{2}}(-1, i\cos 2\theta, 0, 0, i\sin 2\vartheta, 0) . \tag{55}$$

To merge imaginary components $i\cos 2\theta$ and $i\sin 2\theta$ of vectors $\mathbf{A}_{W_2}$ and $\tilde{\mathbf{B}}_{W_2}$, one applies operator $\exp(i\xi/2\sigma_{yy}^{ab})$, $\xi = (\pi/2 - 2\theta)$ which is a 2-5 rotation by angle $\xi$. For the state

$$\left|\mathbf{W}_3\right\rangle = \exp(i\xi/2\sigma_{yy}^{ab})\left|W_2\right\rangle = 1/2\left(\langle 001| + i\langle 010| + \langle 100| + i\langle 111|\right) \tag{56}$$

we have

$$\mathbf{A}_{W_3} = (-1, 0, 0)/2, \quad \mathbf{B}_{W_3} = (0, -1, 0)/2, \quad \mathbf{C}_{W_3} = (-1, 0, 0)/2 \tag{57}$$

Interestingly, transformation $\exp(i\xi/2\sigma_{yy}^{ab})$ efficiently drove parameter $\theta$ in equation (55) to take value $\theta = \pi/4$, as in equation (56), which was expected to happen due to geometric design of the SO(6) transformation on vector $\left(\mathbf{A}, \tilde{\mathbf{B}}\right)_{W_2}$. As we have discussed above, the value $\theta = \pi/4$ is a special case: it means that already after first coupling transformation the state $\mathbf{W}_2$ will be locally equivalent to GHZ.

As it clearly follows from (57), entanglement parameters of $\mathbf{W}_3$ state match those of **GHZ** state. To design next transformation we remind that for GHZ state (47a) we have $\mathbf{A}_{GHZ}, \mathbf{B}_{GHZ}, \mathbf{C}_{GHZ} = (0, 0, 1)/2$ which apparently differs from (57). This difference can be corrected by a set of local transformations: $\exp(i\pi/4\sigma_x^b) \Box \exp(\pi/2l_{23}^b)$, $\exp(-i\pi/4\sigma_y^c) \Box \exp(\pi/2l_{31}^c)$, $\exp(-i\pi/4\sigma_y^a) \Box \exp(\pi/2l_{31}^a)$. These operators will transfer $-1/2$ entries of $\mathbf{A}_{W_3}$ $\mathbf{B}_{W_3}$ $\mathbf{C}_{W_3}$ to the third position, as in $\mathbf{A}_{GHZ}$, $\mathbf{B}_{GHZ}$, $\mathbf{C}_{GHZ}$. After these transformations we have

$$\left|\mathbf{W}_4\right\rangle = e^{-i\pi/4\sigma_y^a} e^{-i\pi/4\sigma_y^c} e^{i\pi/4\sigma_x^b} \left|\mathbf{W}_3\right\rangle = 1/\sqrt{2}\left(-\langle 000| + i\langle 111|\right) \tag{58}$$

$$\mathbf{A}_{W_4} = (0, 0, 1)/2, \quad \mathbf{B}_{W_3} = (0, 0, 1)/2, \quad \mathbf{C}_{W_3} = (0, 0, 1)/2 \tag{59}$$

State $\left|\mathbf{W}_4\right\rangle$ just needs one local $\exp(-i\pi/4\sigma_z^a)$ rotation and multiplication by factor (-1) to match the target state (47):

$$\left|\mathbf{W}_5\right\rangle = (-1)e^{-i\pi/4\sigma_z^a}\left|\mathbf{W}_4\right\rangle = 1/\sqrt{2}\, e^{-i\pi/4}\left(|0,0,0\rangle + |1,1,1\rangle\right)$$

Interestingly, local operation $\exp(-i\pi/4\sigma_z^a) \Box \exp(-\pi/2l_{21}^a)$, as well as multiplication by (-1), has no effect on (50d) due to zero entries in positions 1,2.

Overall for $|W\rangle \to |GHZ\rangle$ transformation we have a sequence of the following operators:

$$\mathbf{U}(\mathbf{W} \to \mathbf{GHZ}) = (-1)e^{-i\pi/4\sigma_z^a} e^{-i\pi/4\sigma_y^a} e^{-i\pi/4\sigma_y^c} e^{i\pi/4\sigma_x^b} e^{i(\pi/4-\theta)\sigma_{yy}^{ab}} e^{i\varphi/2\sigma_z^c} e^{-i\varphi/2\sigma_z^b} e^{i\pi/4\sigma_{xx}^{bc}} \tag{60}$$

Expression (60) seems to be a rather long sequence of operators. However, most of these operations are local rotations: there are only two coupling operations and six local rotations. Apparently, identification of the series of local rotations required to implement the whole sequence was an important part of the general design, and as we have seen above this part has very simple character. Due to **ABC** - vectors representation of tree-qubit states each step in this sequence was designed following simple logic of SO(3) rotations. Also, notice that we have not applied any rotations outside of selected two-dimensional coordinate planes for local as well as coupling transformations. In some sense all of these transformations are a sequence of simple SO(2) rotations.

## XI. GEOMETRIC PROPERTIES OF THREE-TANGLE.

### 1. Three-Tangle Extremum.

The problem we are going to solve first using geometry of **ABC** - vectors has the following physical meaning: given an arbitrary three-qubit state, what is the maximal value of three-tangle achievable only by two-qubit coupling? It other words, there are three parties: Alice Bob and Charlie. They share a three-qubit entangled state. Alice and Bob can implement any operation on the $a-b$ pair of qubits, but they do not have access to qubit $c$. What is the maximum value of three tangle which can be achieved in this case?

First we find (local) extrema of three-tangle using SO(6) geometry by differentiating three tangle $\tau_{abc} = 4|\mathbf{A}^2|$ over all possible variations of $(\mathbf{A},\tilde{\mathbf{B}})$ vector under SO(6) action. Since evolution of $(\mathbf{A},\tilde{\mathbf{B}})$ is rendered by some $\Upsilon \subset SO(6)$ operator, infinitesimal variation of $(\mathbf{A},\tilde{\mathbf{B}})$ is given by so(6) tangent operator $d\Upsilon = \sum_{n,m} d\xi_{n,m} \chi^{(n,m)}$, $\chi^{(n,m)} \subset so(6)$ acting on $(\mathbf{A},\tilde{\mathbf{B}})$, i.e. $d(\mathbf{A},\tilde{\mathbf{B}}) = d\xi_{n,m}\chi^{(n,m)}(\mathbf{A},\tilde{\mathbf{B}})$, where one can have so(6) algebra spanned by antisymmetric $6\times 6$ matrices with matrix elements $\left[\chi^{(n,m)}\right]_{i,j} = -\delta_{i,n}\delta_{m,j} + \delta_{i,m}\delta_{n,j}$, for example. Since three-tangle $\tau_{abc}$ does not change under the action of $SO(3)^{(a)} \otimes SO(3)^{(b)}$ subgroup of local transformations we need to consider only those generators $\chi^{(n,m)}$ which mix vectors $\mathbf{A}$ and $\tilde{\mathbf{B}}$, namely $\Lambda_{n,m}^{(ab)} \equiv \chi^{(n,m+3)}$ as defined in equation (18), such that

$$d\mathbf{A}_i = \sum_{k=1,2,3} d\xi_{i,k+3} \tilde{\mathbf{B}}_k \equiv -i \sum_{k=1,2,3} d\xi_{i,k+3} \mathbf{B}_k \qquad (61)$$

The differential $d\tau_{abc} = -1/\tau_{abc} \operatorname{Im}\left(\mathbf{A}^{2*} \sum_{i,j} A_i B_j d\xi_{i,j+1}\right)$ such that equation for extremum $d\tau_{abc} = 0$ becomes

$$\operatorname{Im}\left(\mathbf{A}^{2*} A_i B_j d\xi_{i,j}\right) = 0 \qquad (62a)$$

To solve equation (62) we introduce phases

$$\Phi^{(A)} = 1/2 \operatorname{Arg}(\mathbf{A}^2) \qquad (62b)$$

$$\varphi_i^{(A)} = \operatorname{Arg}(|A_i|), \ \varphi_i^{(B)} = \operatorname{Arg}(|B_i|), i=1,2,3 \qquad (62c)$$

such that $\mathbf{A}^2 = |\mathbf{A}^2| e^{2i\Phi^{(A)}}$ and $A_i = |A_i| e^{i\varphi_i^{(A)}}$, $B_i = |B_i| e^{i\varphi_i^{(B)}}$. Since $d\xi_{j,i}$ are free parameters, equation (62a) is equivalent to $|A_i||B_j| \operatorname{Im} e^{2i\Phi^{(A)} + i\varphi_i^{(a)} + i\varphi_j^{(b)}} = 0$, $i,j = 1,2,3$, or

$$\varphi_i^{(A)} + \varphi_j^{(B)} - 2\Phi^{(A)} = 0, \ i,j = 1,2,3 \qquad (63)$$

Solution of equation (63) is

$$\varphi_{1,2,3}^{(A)} = \varphi_{1,2,3}^{(B)} = \Phi^{(A)} \qquad (64)$$

At the extremum of $\tau_{abc}$ all components of vectors $\mathbf{A}$ and $\mathbf{B}$ have identical phases and can be made real by multiplication by single phase factor $\exp(-i\Phi^{(A)})$, which is equivalent to multiplication of $|\psi\rangle$ by $\exp(-i/2\,\Phi^{(A)})$. Another observation which follows from this derivation is that extremum of $\tau_{abc}$ is achievable by some SO(6) transformation since this group is a connected compact manifold. As we illustrate below there is only one local maximum $\tau_{abc}$ which is also a global maximum of.

## 2. Three-Tangle Geometry

We start with an observation that according to equation (27) the sum of magnitudes of six components of vector $(\mathbf{A},\mathbf{B})$ is $SU(4)^{(ab)}$ invariant (as well as $SO(6)^{ab}$ invariant)

$$\sum_{n=1}^{3}\left(\left|A_{n}\right|^{2}+\left|B_{n}\right|^{2}\right)=1/2\,\tau_{c(ab)} \tag{65}$$

This sum does not change under the a-b coupling. We would like to rewrite expression (10) for three-tangle in the form algebraically similar to equation (65). Due to Plücker relation $\mathbf{A}^2=\mathbf{B}^2$, equation (9), three-tangle can be expressed as

$$1/2\,\tau_{abc}=\left|\mathbf{A}^{2}+\mathbf{B}^{2}\right|\equiv\left|\sum_{n=1}^{3}(A_{n}^{\,2}+B_{n}^{\,2})\right| \tag{66}$$

In other words, $\tau_{abc}/2$ is equal to the magnitude of the sum $A_1^2+A_2^2+A_3^2+B_1^{\,2}+B_2^{\,2}+B_3^{\,2}$ while $\tau_{c(ab)}/2$ is the sum of magnitudes of the same set of numbers $\left|A_1^2\right|+\left|A_2^2\right|+\left|A_3^2\right|+\left|B_1^{\,2}\right|+\left|B_2^{\,2}\right|+\left|B_3^{\,2}\right|$.

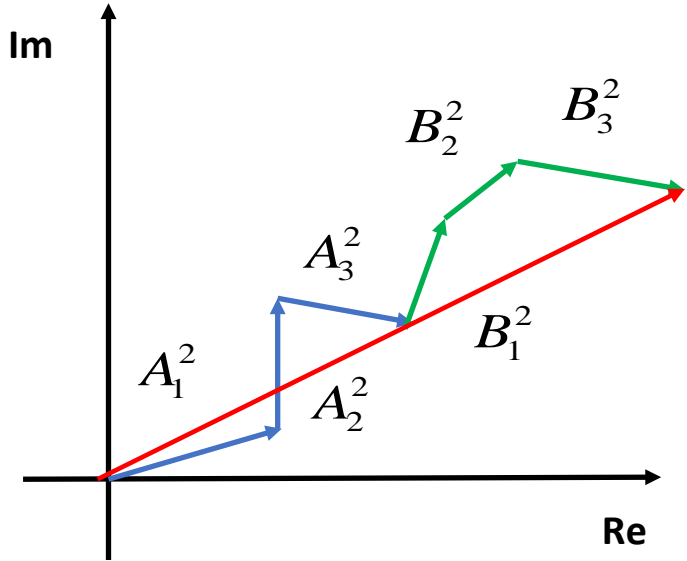

Fig 1. A complex-plane diagram illustrating the geometric nature of the inequality $\tau_{c(ab)}\geq\tau_{abc}$.

Equation (64) is stipulating that maximum of the sum $\sum_{n=1}^{3}\left(\left|A_{n}\right|^{2}+\left|B_{n}\right|^{2}\right)=1/2\,\tau_{c(ab)}$, occurs when $\arg(A_{1,2,3})=\arg(B_{1,2,3})=const$, meaning that all arrows representing six complex numbers on the diagram above will be strictly aligned. For example, all six numbers are real: $\mathrm{Im}(A_{1,2,3})=\mathrm{Im}(B_{1,2,3})=0$. Next trivial but important fact is that $\left|(\mathbf{A},\mathbf{B})\cdot(\mathbf{A},\mathbf{B})\right|\leq(\mathbf{A},\mathbf{B})\cdot(\mathbf{A},\mathbf{B})^{*}$ and while components of vector $(\mathbf{A},\mathbf{B})$ change under SO(6) action the Hermitian inner product $(\mathbf{A},\mathbf{B})\cdot(\mathbf{A},\mathbf{B})^{*}$ is an invariant.

A naïve way of thinking about equation (66) is to imagine that all arrows representing complex numbers $A_1^2, A_2^2$ etc. on the Fig.1 can be rotated by a phase multiplication to get perfectly aligned with each other. Such a concept is incorrect: wave function can be multiplied only by a single global phase. One does not have independent control over phases of all individual components of vectors $\mathbf{A}$ and $\mathbf{B}$. Control operations include a) global phase multiplication, b) $SO(3)^{a,b}$ rotations and c) $SO(6)$ coupling operations. According to (65) and (66) the upper bound for $\tau_{abc}$ is $\tau_{c(ab)}$ and it turns out that this bound can be achieved by a set of local transformations and single qubit-qubit coupling $\exp(\lambda i \sigma_{n,m}^{(ab)})$ such that

$$\operatorname*{Max}_{SU(4)^{ab}}(\tau_{abc}) = \tau_{c(ab)}, \tag{67}$$

Note that equation (64) implies that two tangles (11a,b,c) are zero when $\tau_{abc}$ reaches its maximal value. Since $\tau_{abc} = \tau_{c(ab)} - \tau_{(ac)} - \tau_{(bc)}$, bipartite entanglement between qubits $a$ and $c$, and $b$ and $c$, is converted into the three-tangle form. Conversely, three-tangle can be transformed into a two-qubit entanglement. Below, we explore how this can be constructively achieved.

## XII. GAUGE REDUCTION OF ENTANGLEMENT EQUATIONS

Our next goal is to understand which operations can be applied to achieve the maximum (67). Let us first clarify how global phase multiplication affects 6D vector $(\mathbf{A}, \tilde{\mathbf{B}})$ and 3D vectors $A, B, C$. We will be using it below for computing optimal coupling operations. Strictly speaking, the set of $SO(6)$ operators does not include multiplication by a phase. However, one can formally enlarge $SU(4)$ group to $U(4) = SU(4) \times U(1)$ such that $SO(6)$ also expands to include phase multiplication $SO(6) \times U(1) \cong U(4)/\mathbb{Z}_2$. The image of $e^{i\varphi} \in U(4)$ in $SO(6) \times U(1)$ group is $e^{2i\varphi}$. Multiplication by phase commutes with both $SU(4)$ and $SO(6)$ operations such that $U(1)$ group does not increase the algebraic complexity of these groups.

To simplify equations, we will denote vector $(\mathbf{A}, \tilde{\mathbf{B}})$ as $\mathbf{Q}$. This vector in general has real and imaginary parts $\mathbf{Q} = \mathbf{Q}_R + i\mathbf{Q}_I$. Since Plücker relation guarantees that $\mathbf{Q}^2 = (\mathbf{Q}_R)^2 - (\mathbf{Q}_I)^2 + 2i\mathbf{Q}_R \cdot \mathbf{Q}_I = 0$ we have

$$\mathbf{Q}_R^2 = \mathbf{Q}_I^2, \quad \mathbf{Q}_R \mathbf{Q}_I = 0 \tag{68}$$

Multiplication by global phase $\exp(i\phi)$ will mix imaginary and real parts of $\mathbf{Q}$:

$$\begin{pmatrix} \mathbf{Q}_R \\ \mathbf{Q}_I \end{pmatrix} \to \begin{pmatrix} \cos 2\phi & -\sin 2\phi \\ \sin 2\phi & \cos 2\phi \end{pmatrix} \begin{pmatrix} \mathbf{Q}_R \\ \mathbf{Q}_I \end{pmatrix} \tag{69}$$

Due to relations (68), $\mathbf{Q}_R^2$, $\mathbf{Q}_I^2$ and $\mathbf{Q}_R \mathbf{Q}_I$ do not change.

Global phase (gauge) does not affect physical observables. However, there is a special gauge which makes algebraic relations between vectors $\mathbf{A}, \mathbf{B}, \mathbf{C}$ and entanglement parameters particularly simple. Multiplication of $|\psi\rangle$ by $\exp(-i/2\, \Phi^{(A)})$, as defined by equation (62b), induces the map $\mathbf{A} \to \mathbf{A} = \mathbf{A}e^{-i\Phi^{(a)}}$. In this case $\mathbf{A}_R = \operatorname{Re}(\mathbf{A})$ and $\mathbf{A}_I = \operatorname{Im}(\mathbf{A})$ are orthogonal, $\mathbf{A}_R \mathbf{A}_I = 0$, because $\mathbf{A}^2 = \mathbf{A}_R^2 - \mathbf{A}_I^2 + 2i\mathbf{A}_R \mathbf{A}_I$ is purely real. We also assume that the gauge is chosen such that $\mathbf{A}^2 \geq 0$, i.e. $\mathbf{A}_R^2 \geq \mathbf{A}_I^2$. Plücker relation (9) also guarantees that $\Phi^{(A)} = \Phi^{(B)} = \Phi^{(C)}$, such that $\mathbf{B}_R \mathbf{B}_I = \mathbf{C}_R \mathbf{C}_I = 0$, $\mathbf{B}_R^2 \geq \mathbf{B}_I^2$ and $\mathbf{C}_R^2 \geq \mathbf{C}_I^2$.

In this gauge entanglement parameters simplify as

$$\begin{aligned} \tau_{abc} &= 4(\mathbf{A}_R^2 - \mathbf{A}_I^2) = 4(\mathbf{B}_R^2 - \mathbf{B}_I^2) = 4(\mathbf{C}_R^2 - \mathbf{C}_I^2) \\ \tau_{(bc)} &= 4\mathbf{A}_I^2, \ \tau_{(ac)} = 4\mathbf{B}_I^2, \ \tau_{(ab)} = 4\mathbf{C}_I^2 \end{aligned} \tag{70}$$

# XIII. EXAMPLES OF SO(6) TRANSFORMATIONS MAXIMIZING THREE-TANGLE.

As an illustration of vector control technique, we first describe a simple, but not necessarily an optimal transformation. Consider a random state $|\psi\rangle$ which generates a set of three complex vectors $\mathbf{A}$, $\mathbf{B}$, $\mathbf{C}$ and a set of associated entanglement parameters. Suppose that one can perform coupling operations only on two qubits, which we assume to be qubits $a$ and $b$. This restricts the possibility to have full control of three-qubit system. The goal is to perform a series of local operations followed by qubit couplings such as to maximize the three-tangle as in equation (67). For simplicity we assume that we chose special gauge (62b). To further simplify the logic of control operations we perform local rotations such that $\mathbf{A}_R$ is aligned with the first axis and has zero components $A_{R,2} = A_{R,3} = 0$. Then a local rotation is performed to align vector $\mathbf{A}_I$ with the third axis, for example, which can be completed because $\mathbf{A}_R \perp \mathbf{A}_I$. A similar set of local transformations is performed on vectors $\mathbf{B}_R$ and $\mathbf{B}_I$ such that before we apply coupling operations, which mix vectors $\mathbf{A}$ and $\mathbf{B}$, we have

$$\mathbf{A} = (A_R, 0, iA_I), \mathbf{B} = (B_R, 0, iB_I) \tag{71}$$

Equations (70) reduce to

$$\tau_{abc} = 4(A_R^2 - A_I^2) = 4(B_R^2 - B_I^2)$$
$$\tau_{(bc)} = 4A_I^2, \tau_{(ac)} = 4B_I^2 \tag{72}$$
$$\tau_{c(ab)} = 2\left(A_R^2 + A_I^2 + B_R^2 + B_I^2\right)$$

To identify a required qubit coupling we need to find how to couple components 1-2-3 and 4-5-6 of 6D vector $\mathbf{Q} = (A_R, 0, iA_I, -iB_R, 0, B_I)$. There are two options. One can perform a sequence of two operations: $\pi/2$ rotation in the 2-6 plane, and second $\pi/2$ rotation in the 3-5 plane. Both rotations are physically implemented by unitary operators $\exp(i\pi/4\sigma_{yz}^{(ab)})$, $\exp(i\pi/4\sigma_{zy}^{(ab)})$ correspondingly. These operators are locally equivalent to CZ gate. The result of these actions is

$$\mathbf{Q}^{(\max)} = (A_R, B_I, 0, -iB_R, iA_I, 0) \tag{73}$$

$$\mathbf{A}_R^{(\max)} = (A_R, 0, B_I), \mathbf{A}_I^{(\max)} = (0,0,0)$$
$$\mathbf{B}_R^{(\max)} = (B_R, 0, A_I), \mathbf{B}_I^{(\max)} = (0,0,0) \tag{74}$$

Evidently vectors (74) satisfy the extremum condition (64) since $\varphi_{1,2,3}^{(a)} = \varphi_{1,2,3}^{(b)} = 0$. Let us demonstrate that maximum value (67) is achieved by these transformation. According to (72), (74) and equation (9) we have three-tangle

$$\tau_{abc}^{(\max)} = 4(A_R^2 + B_I^2) \equiv 4(B_R^2 + A_I^2). \tag{75}$$

At the same time $\tau_{c(ab)} = 2(A_R^2 + B_I^2 + B_R^2 + A_I^2)$ such that (67) is satisfied.

However, there is a shorter way to achieve the same goal by applying only one coupling operation: $\pi/2$ rotation in the 3-6 plane implemented by $\exp(i\pi/4\sigma_{zz}^{(ab)})$ resulting in $\mathbf{Q}^{(\max)} = (A_R, 0, B_I, -iB_R, 0, -iA_I)$.

More economical two-coupling scheme can be implemented by rotation in the plane 1-6 by the angle $\theta_{1,6} = \arg(A_R + iB_I)$ followed by 3-4 rotation $\theta_{3,4} = \arg(B_R + iA_I)$, resulting in

$$\mathbf{Q}^{(\max)} = \left(\sqrt{A_R^2 + B_I^2}, 0, 0, -i\sqrt{A_I^2 + B_R^2}, 0, 0\right) \tag{76}$$

such that $\tau_{abc}^{(\max)} = 4(A_R^2 + B_I^2) \equiv 4(B_R^2 + A_I^2)$ as in equation (75). Notice that if $A_R > B_I$ and $B_R > A_I$ both angles $\theta_{1,6}$ and $\theta_{3,4}$ can be made smaller than $\pi/4$ which is two times faster than two $\pi/2$ rotations discussed in the first example, and can also be faster than a single $\pi/2$ rotation described in the second example. Physically in both cases the resource for maximizing three-tangle are vectors $\mathbf{A}_I$ and $\mathbf{B}_I$, which define two-tangles via equation (72).

## XIV. ENTANGLEMENT CONTROL BY $\mathrm{USp}(4) \cong \mathrm{Spin}(5)$ QUATERNIONIC SUBGROUP OF $\mathrm{SU}(4)$

### 1. Quaternionic states and quaternionic operators.

There are two types of problems in quantum control theory: control of quantum states and control of quantum transformation. By state control we mean a set of problems where the goal is to transform given initial state $\Psi$ into some final state $\Phi$. The goal of control on transformations, or gates, is to generate specific transformation (in general not necessarily a unitary one). The difference is that (unitary) transformations act on the entire Hilbert space by acting on every state from a complete basis set while state control implies that transformation is limited to the action on a specific state. Apparently, if there are enough resources for complete gate control then it allows to establish control of state transformations. However, in general one does not need a complete control over transformations to be able to control states [35]. This feature has been exploited in our previous work on fusion gate optimization where we considered operator action on 2-dimensional subspace of four-dimensional Hilbert space [36]). Limited gate control mathematically implies that Hamiltonian includes only a subset of complete su($n$) or sl($n$) Lie algebra such that this subset closes under commutation in one of subalgebras of the full Lie algebra. Typical example, which we will be studying here is $\mathrm{USp}(4)$ subgroup of $\mathrm{SU}(4)$ which has been also studied in [37, 38].

A straightforward way to set up the $\mathrm{USp}(4)$ subgroup of $\mathrm{SU}(4)$ is to arrange a $2\times 2$ quaternionic matrix with entries $\mathbf{x}_{nm} \subset \mathbb{H}$, $m,n=1,2$ satisfying unitarity condition $\sum \mathbf{x}_{nm}\mathbf{x}^\dagger_{nk} = \boldsymbol{\delta}_{mk}$. To fix notation, we use boldface for quaternions, italic boldface is used for 4-dimensional vector $\boldsymbol{x} = (x_0, x_1, x_2, x_4) \in \mathbb{R}^4$, and vector part of $\boldsymbol{x}$ is denoted as $\vec{x} = (x_1, x_2, x_3)$. For example, quaternion $\mathbf{y} = y_0\boldsymbol{1} + y_1\boldsymbol{i} + y_2\boldsymbol{j} + y_3\boldsymbol{k}$. Pauli-matrix representation for a quaternion $\mathbf{x}$ has a $2\times 2$ matrix form $\mathbf{x} = x_0 \boldsymbol{I}_{(2\times 2)} - i\vec{x}\vec{\sigma}$. In this representation $2\times 2$ matrix $\mathbf{x}_{nm}$ becomes 4x4 complex-valued matrix. Minus sign in this equation is required to match the quaternionic multiplication rule $\boldsymbol{i}\times \boldsymbol{j} = \boldsymbol{k}$, which takes the form $(-i\sigma_x)\times(-i\sigma_y) = (-i\sigma_z)$. For $\mathrm{USp}(4)$ unitary we have $x_0^2 + x_1^2 + x_2^2 + x_3^2 = 1$ which is an $S^3$ sphere embedded in $\mathbb{R}^4$.

Importantly, we can easily identify the subset of 10 elements of su(4) Lie algebra which are generating the entire $\mathrm{USp}(4)$ group. For $a(bc)$ partition

$$\{i\sigma_y^{(b)}\}, \{i\sigma_{xx}^{(bc)}, i\sigma_{zx}^{(bc)}\}, \{\sigma_{xy}^{(bc)}, \sigma_{zy}^{(bc)}, \sigma_z^{(c)}\}, \{\sigma_{xz}^{(bc)}, \sigma_{zz}^{(bc)}, \sigma_y^{(c)}, \sigma_x^{(c)}\}. \tag{77}$$

Generators are arranged by $\mathrm{Cliff}_n$ $n=1,2,3,4$ algebra sets. Each entry in (77) represents a $2\times 2$ quaternionic matrix. Generator $i\sigma_z^{(b)}$, for example, is not in this list because it includes two diagonal entries $i\mathbf{I}_{(2\times 2)}$ which do not belong to the algebra of quaternions represented by $\mathrm{Span}\{\boldsymbol{I}_{(2\times 2)}, i\sigma_{1,2,3}\}$. Corresponding set of so(6) generators is

$$\{l_{13}^b\}, \{\Lambda_{11}^{(bc)}, \Lambda_{31}^{(bc)}\}, \{\Lambda_{12}^{(bc)}, \Lambda_{32}^{(bc)}, l_3^{(c)}\}, \{\Lambda_{13}^{(bc)}, \Lambda_{33}^{(bc)}, l_{13}^{(c)}, l_{23}^{(c)}\}. \tag{78}$$

Strictly speaking, $\mathrm{SU}(4)$ group is isomorphic to $\mathrm{Spin}(6)$ Lie group (a universal double cover of $\mathrm{SO}(6)$). The set in equation (77) is generating $\mathrm{Spin}(5)$ subgroup of $\mathrm{Spin}(6)$ which is a double cover of $\mathrm{SO}(5)$ subgroup of $\mathrm{SO}(6)$. Correspondingly, the set (78) is generating $\mathrm{SO}(5)$ rotations acting on components (1,3,4,5,6) of vector $(\mathbf{B}, \tilde{\mathbf{C}})$. As one can see the action of these operators on vector $(\mathbf{B}, \tilde{\mathbf{C}})$ is limited to those couplings which do not affect second component of this vector. Generators $(l_{12}^b, l_{32}^b, \Lambda_{21}^{(bc)}, \Lambda_{22}^{(bc)}, \Lambda_{23}^{(bc)})$ corresponding to $\{i\sigma_z^{(b)}, i\sigma_x^{(b)}, i\sigma_{yx}^{(bc)}, i\sigma_{yy}^{(bc)}, i\sigma_{yz}^{(bc)}\}$ are eliminated from the full list of SO(6) generators. For example, generator $i\sigma_{yy}^{(bc)} = i\sigma_y^{(b)} \otimes \sigma_y^{(c)} = \begin{pmatrix} 0 & \sigma_y \\ -\sigma_y & 0 \end{pmatrix}$ does not represent a quaternionic matrix.

After we set up the restriction (77) on generators, we also impose a similar restriction on the space of possible states. We include only those states which satisfy the following condition: state coefficients arranged in the form of $2\times 4$ matrix corresponding to partition $a(bc)$, equation (2a), are also required to be quaternionic, i.e. this matrix is represented as an $\mathbb{H}^2$ vector

$$\begin{pmatrix} c_{000} & c_{100} \\ c_{001} & c_{101} \\ c_{010} & c_{110} \\ c_{011} & c_{111} \end{pmatrix} = \begin{pmatrix} \mathbf{x} \\ \mathbf{y} \end{pmatrix} \equiv \begin{pmatrix} x_0 + ix_3 & ix_1 + x_2 \\ ix_1 - x_2 & x_0 - ix_3 \\ y_0 + iy_3 & iy_1 + y_2 \\ iy_1 - y_2 & y_0 - iy_3 \end{pmatrix} \qquad (79)$$

Since $su(2)^{(a)}$ set $\{i\sigma_{x,y,z}^{(a)}\}$ is consistent with quaternionic state (79), the action of local operators $SU(2)^{(a)}$, implemented by matrix right-multiplication, should be also included in the set of control transformations on state (79), i.e. we have $usp(4)^{(bc)} \oplus su(2)^{(a)}$ Lie algebra of quaternionic generators. Notice that we have only one generator for local transformations on qubit $b$ and the full set of local transformations of qubits $c$ and $a$. The property (79) of the state $c_{i,j,k}$ is partition-specific: for partition $b(ca)$ matrix (2b) does not in general satisfy equation (79) with real-values coefficients $x_n$ and $y_n$. However, this form is consistent with $a \leftrightarrow c$ permutation. Such a permutation of matrix (79) generates a quaternionic matrix with transposed quaternionic entries:

$$\begin{pmatrix} c_{000} & c_{001} \\ c_{100} & c_{101} \\ c_{010} & c_{011} \\ c_{110} & c_{111} \end{pmatrix} = \begin{pmatrix} \mathbf{x}^T \\ \mathbf{y}^T \end{pmatrix}. \qquad (80)$$

As we will see below vector $\mathbf{A}$ can be obtained from vector $\mathbf{C}$ by replacing $\mathbf{x}$ and $\mathbf{y}$ with $\mathbf{x}^T$ and $\mathbf{y}^T$ correspondingly.

The state (79) is determined by a set of eight real coefficients $x_{0,1,2,3}$ and $y_{0,1,2,3}$. This is a $R^8 \equiv R^4 \oplus R^4$ subspace of full $\mathbb{C}^8 \equiv \mathbb{R}^{16}$ space of quantum states. One can show that $\langle\psi|\psi\rangle = 2(\mathbf{xx}^\dagger + \mathbf{yy}^\dagger) = 2(\mathbf{x}\cdot\mathbf{x} + \mathbf{y}\cdot\mathbf{y})$, such that for normalized sates $\mathbf{x}\cdot\mathbf{x} + \mathbf{y}\cdot\mathbf{y} = 1/2$, and the set of normalized states (79) represents an $S^7$ sphere embedded in the $S^{15}$ sphere.

Quaternionic $USp(4)$ operators which *are not* coupling $\mathbf{x}$- and $\mathbf{y}$-entries in equation (79) (do not confuse this with qubit-qubit coupling) are limited to six operators forming Spin(4) group $\{\sigma_z^{(c)}\}, \{\sigma_y^{(c)}, \sigma_x^{(c)}\}, \{\sigma_{zx}^{(bc)}, \sigma_{zy}^{(bc)}\sigma_{zz}^{(bc)}\}$, plus the $su(2)^{(a)}$ set $\{i\sigma_{x,y,z}^{(a)}\}$. The $bc$ set corresponds to SO(4) subgroup of SO(6) generated by $\{l_{12}^{(c)}, l_{13}^{(c)}, l_{23}^{(c)}, \Lambda_{31}^{(bc)}, \Lambda_{32}^{(bc)}, \Lambda_{33}^{(bc)}\}$. These are rotations of 3-4-5-6 components of vector $(\mathbf{B}, \tilde{\mathbf{C}})$. Importantly, local operations on qubit $a$ are consistent with (79) structure because any local $SU(2)^{(a)}$ operator $\mathbf{U}^{(a)} = \xi_0^a I + \sum_{n=1}^{3} \xi_n^a i\sigma_n^a$ is acting as right-side multiplication of the state in equation (79):

$$\begin{pmatrix} \mathbf{x} \\ \mathbf{y} \end{pmatrix} \to \begin{pmatrix} \mathbf{x}\xi^a \\ \mathbf{y}\xi^a \end{pmatrix}, \quad \xi^a = I_{2\times 2}\xi_0^a - \sum_{n=1}^{3} \xi_n^a i\sigma_n \qquad (81)$$

Let's summarize control properties of these generators: local operations are generated by $\{i\sigma_y^{(b)}\}$, $\{\sigma_z^{(c)}, \sigma_y^{(c)}, \sigma_x^{(c)}\}$ and $\{\sigma_z^{(a)}, \sigma_y^{(a)}, \sigma_x^{(a)}\}$; the set of six coupling generators is $\{\sigma_{xx}^{(bc)}, \sigma_{zx}^{(bc)}, \sigma_{xy}^{(bc)}, \sigma_{zy}^{(bc)}, \sigma_{xz}^{(bc)}, \sigma_{zz}^{(bc)}\}$. The su(2) operators on **a** and **c** qubits are not mixing **x** and **y** entries in equation (79).

There are two special sets consisting of three operators in each set:

$$\frac{1}{2}(\sigma_z^{(c)} \pm \sigma_{zx}^{(bc)}) \,,\; \frac{1}{2}(\sigma_z^{(c)} \pm \sigma_{zy}^{(bc)}),\; \frac{1}{2}(\sigma_z^{(c)} \pm \sigma_{zy}^{(bc)}), \tag{82}$$

These operators can be represented as block-diagonal matrices embedded in $4\times 4$ matrix. For example

$$\frac{1}{2}(\sigma_z^{(c)} + \sigma_{z(x,y,z)}^{(bc)}) = \begin{pmatrix} \sigma_{x,y,z} & 0 \\ 0 & 0 \end{pmatrix}$$
$$\frac{1}{2}(\sigma_z^{(c)} - \sigma_{z(x,y,z)}^{(bc)}) = \begin{pmatrix} 0 & 0 \\ 0 & \sigma_{x,y,z} \end{pmatrix} \tag{83}$$

Under an appropriate similarity transform there operators take the form of $s$ and $\tau$ operators introduced in [39]. Generators (83) provide independent control over **x** and **y** quaternions ($x$ and $y$ vectors correspondingly). We remind that Euler's formula $\exp(i\gamma\vec{n}\vec{\sigma}) = \cos\gamma \mathbf{I} + i\sin\gamma \vec{n}\vec{\sigma}$ allows one to generate an arbitrary unit quaternion, such that

$$\exp\left[\begin{pmatrix} i\gamma_x \vec{n}_x \vec{\sigma} & 0 \\ 0 & i\gamma_y \vec{n}_y \vec{\sigma} \end{pmatrix}\right] = \begin{pmatrix} \mathbf{s}_x & 0 \\ 0 & \mathbf{s}_y \end{pmatrix} \tag{84}$$

where $\mathbf{s}_{x,y} = \exp(i\gamma_{x,y}\vec{n}_{x,y}\vec{\sigma})$ are unit quaternions. The action of this operator on **x** and **y** is a quaternionic multiplication $\mathbf{x}_{in} \to \mathbf{x}_{out} = \mathbf{s}_x \mathbf{x}_{in}$, $\mathbf{y}_{in} \to \mathbf{y}_{out} = \mathbf{s}_y \mathbf{y}_{in}$. Since $\mathbf{s}_{x,y}$ are unit quaternions this multiplication does not change the norm: $\mathbf{x}_{out}^\dagger \mathbf{x}_{out} = \mathbf{x}_{in}^\dagger \mathbf{s}_x^\dagger \mathbf{s}_x \mathbf{x}_{in} = \mathbf{x}_{in}^\dagger \mathbf{x}_{in}$. Quaternions represent a division algebra, therefore this action is transitive on the set of vectors of equal norm i.e. as long as $\mathbf{x}_{out}^\dagger \mathbf{x}_{out} = \mathbf{x}_{in}^\dagger \mathbf{x}_{in}$ there exists $\mathbf{s}_x$ such that $\mathbf{x}_{out} = \mathbf{s}_x \mathbf{x}_{in}$ (the action is also free since $\mathbf{x}_{out} = \mathbf{x}_{in}$ means $\mathbf{s}_x = \mathbf{I}$). Apparently a desired transformation will be achieved by assigning $\mathbf{s}_x = \mathbf{x}_{out}\mathbf{x}_{in}^{-1}$, $\mathbf{s}_y = \mathbf{y}_{out}\mathbf{y}_{in}^{-1}$ in equation (84).

$$\begin{pmatrix} \mathbf{s}_x & 0 \\ 0 & \mathbf{s}_y \end{pmatrix} \begin{pmatrix} \mathbf{x}_{in} \\ \mathbf{y}_{in} \end{pmatrix} = \begin{pmatrix} \mathbf{x}_{out}\mathbf{x}_{in}^{-1}\mathbf{x}_{in} \\ \mathbf{y}_{out}\mathbf{y}_{in}^{-1}\mathbf{y}_{in} \end{pmatrix} = \begin{pmatrix} \mathbf{x}_{out} \\ \mathbf{y}_{out} \end{pmatrix} \tag{85}$$

It is easy to verify that $|\mathbf{s}_{x,y}| = |\mathbf{x}_{out}||\mathbf{x}_{in}|^{-1} = |\mathbf{y}_{out}||\mathbf{y}_{in}|^{-1} = 1$.

Restriction of invariance of the norm for these transformations is in fact mitigated by operators $i\sigma_y^{(b)}$. The action of this operator is rather unusual. This is the only generator from the set of local transformations on qubit **b** which belongs to the $USp(4)$ group. Its action is generated by the following U(1) group:

$$\exp[i\chi\sigma_y^{(b)}] = \exp\left[i\chi\begin{pmatrix} 0 & -\mathbf{I}_{2\times 2} \\ \mathbf{I}_{2\times 2} & 0 \end{pmatrix}\right] = \begin{pmatrix} \cos\chi & -\sin\chi \\ \sin\chi & \cos\chi \end{pmatrix} \otimes \mathbf{I}_{2\times 2}. \tag{86}$$

Its action on the state (79) is

$$\begin{pmatrix} \mathbf{x}_{out} \\ \mathbf{y}_{out} \end{pmatrix} = \begin{pmatrix} \cos\chi & -\sin\chi \\ \sin\chi & \cos\chi \end{pmatrix} \begin{pmatrix} \mathbf{x}_{in} \\ \mathbf{y}_{in} \end{pmatrix}, \tag{87}$$

The invariant of this action, as one may expect from general considerations, is the norm of the state (79) $\mathbf{xx}^\dagger + \mathbf{yy}^\dagger \equiv x^2 + y^2$. Importantly, this action also generates su(2) action on an $\square^2$ vector with components $\Delta = x^2 - y^2$ and $\Omega = 2xy$ (notice that dot product $2xy = \mathbf{xy}^\dagger + \mathbf{yx}^\dagger$). From (87) we have

$$\begin{pmatrix} \Delta_{out} \\ \Omega_{out} \end{pmatrix} = \begin{pmatrix} \cos 2\chi & -\sin 2\chi \\ \sin 2\chi & +\cos 2\chi \end{pmatrix} \begin{pmatrix} \Delta_{in} \\ \Omega_{in} \end{pmatrix} \tag{88}$$

Taking $\chi = 1/2 \arctan(\Delta_{in}/\Omega_{in})$ one can achieve $x_{out}^2 = y_{out}^2$ (in general $\Omega = 0$ does not mean that $y = 0$ or $x = 0$). This relation is important for the transformation of vector $\mathbf{B}$ to a reduced form below.

### 2. Three-Tangle

Now we discuss how vectors $\mathbf{x}$ and $\mathbf{y}$ are related to entanglement properties of three-qubit system and how one can manipulate entanglement via quaternionic representations of $\mathbf{A}$, $\mathbf{B}$, $\mathbf{C}$. Equation (79) results in

$$\begin{aligned}
\mathbf{A} &= 2\left(-x_0 \vec{y}^T + y_0 \vec{x}^T - \vec{x}^T \times \vec{y}^T\right) = \text{VectorPart}(\mathbf{x}^T \mathbf{y}^* - \mathbf{y}^T \mathbf{x}^*) \\
\mathbf{B} &= \left[-i(x^2 - y^2), x^2 + y^2, 2ixy\right] = \left[-i(\mathbf{xx}^\dagger - \mathbf{yy}^\dagger), \mathbf{xx}^\dagger + \mathbf{yy}^\dagger, i(\mathbf{xy}^\dagger + \mathbf{yx}^\dagger)\right]. \\
\mathbf{C} &= 2\left(-x_0 \vec{y} + y_0 \vec{x} - \vec{x} \times \vec{y}\right) = \text{VectorPart}(\mathbf{xy}^\dagger - \mathbf{yx}^\dagger)
\end{aligned} \tag{89}$$

Here $\vec{x}^T \equiv (x_1, -x_2, x_3)$, $\vec{y}^T \equiv (y_1, -y_2, y_3)$. Vector part of a quaternion $\mathbf{q} = q_0 - i\vec{q}\vec{\sigma}$ is defined as $\vec{q} = (q_1, q_2, q_3)$.

Since real and imaginary parts of vectors (89) are orthogonal, we can use the simplified set of equations (72) for entanglement. Taking into account that for a normalized state $x^2 + y^2 = 1/2$, one can parametrize this equation using $|x| = 1/\sqrt{2} \cos\alpha$, $|y| = 1/\sqrt{2} \sin\alpha$, $xy = 1/2 \cos\alpha \sin\alpha \cos\beta$, where $\cos\beta = \hat{x}\hat{y}$ and $\tan\alpha = |y|/|x|$. We have

$$\begin{aligned}
\tau_{abc} &= (\sin\beta \sin 2\alpha)^2 \\
\tau_{(ac)} &= 1 - \tau_{abc} \\
\tau_{a(bc)} &= \tau_{c(ba)} = 1 \\
\tau_{(ab)} &= \tau_{(bc)} = 0
\end{aligned} \tag{90}$$

Maximal entanglement $\tau_{abc} = 1$ apparently is achieved for $x^2 = y^2 = 1/2$ and $xy = 0$ which is consistent with the idea that maximum three-tangle corresponds to zero imaginary part of $\mathbf{B}$.

### 3. Properties of USp(4) Operations.

Consider local transformation. First, we verify that action of local transformations on qubit $a$ does not change $\mathbf{C}$ and $\mathbf{B}$. Such an action is defined in equation (81). For vector $\mathbf{C}$ we have

$$\mathbf{C} \to 2\,\text{VectorPart}(\mathbf{x}\xi_a \xi_a^\dagger \mathbf{y}^\dagger - \mathbf{y}\xi_a \xi_a^\dagger \mathbf{x}^\dagger) \equiv 2\,\text{VectorPart}(\mathbf{xy}^\dagger - \mathbf{yx}^\dagger). \tag{91}$$

Trivially similar calculation confirms that $\mathbf{B}$ also does not change.

Let us consider local transformations on qubit $c$. Such an action has the form of left multiplication

$$\begin{pmatrix} \mathbf{x} \\ \mathbf{y} \end{pmatrix} \to \begin{pmatrix} \xi^c & 0 \\ 0 & \xi^c \end{pmatrix} \begin{pmatrix} \mathbf{x} \\ \mathbf{y} \end{pmatrix} = \begin{pmatrix} \xi^c \mathbf{x} \\ \xi^c \mathbf{y} \end{pmatrix} \tag{92}$$

The induced change in $\mathbf{A}$ is

$$\mathbf{A} \to \text{VectorPart}(\mathbf{x}^T \xi_c^T \xi_c^* \mathbf{y}^* - \mathbf{y}^T \xi_c^T \xi_c^* \mathbf{x}^*) \equiv \text{VectorPart}(\mathbf{x}^T \mathbf{y}^* - \mathbf{y}^T \mathbf{x}^*).$$

Similarly, for vector $\mathbf{B}$ we have trivial transformation of the first two components $\mathbf{xx}^\dagger \to \xi_c \mathbf{xx}^\dagger \xi_c^\dagger$, $\mathbf{yy}^\dagger \to \xi_c \mathbf{yy}^\dagger \xi_c^\dagger$, and two-step calculation for the third component.

$$\mathbf{xy}^\dagger + \mathbf{yx}^\dagger \to \xi_c \left( \mathbf{xy}^\dagger + \mathbf{yx}^\dagger \right) \xi_c^\dagger \equiv \xi_c \xi_c^\dagger \left( \mathbf{xy}^\dagger + \mathbf{yx}^\dagger \right) \equiv \left( \mathbf{xy}^\dagger + \mathbf{yx}^\dagger \right). \tag{93}$$

This is an obvious result: multiplication by a unit quaternion is an element of SO(4) operation acting simultaneously on both vectors $x$ and $y$. Consequently, dot product of these two vectors does not change.

Single-generator local transformation on $\mathbf{B}$, equation (87) also demonstrates that $\mathbf{C}$ and $\mathbf{A}$ do not change, for example,

$$\mathbf{C}_{out} = \mathbf{x}_{out} \mathbf{y}_{out}^\dagger - \mathbf{y}_{out} \mathbf{x}_{out}^\dagger = \mathbf{x}_{in} \mathbf{y}_{in}^\dagger - \mathbf{y}_{in} \mathbf{x}_{in}^\dagger \tag{94}$$

Without going into details of how 3D rotations are rendered by quaternionic multiplication, we just note that an adjoint action by two unit quaternions on a purely imaginary quaternion, i.e. component $q_0 = 0$, is one of the most useful representations of SO(3) group for technical applications. In other words, action (92) results in SO(3) rotations of vector $\mathbf{C}$ while action (81) generates rotations of vector $\mathbf{A}$. And, as it clearly follows from equation (80) local action on qubit $b$ generates 2D rotation of $\mathbf{B}_1$ and $\mathbf{B}_3$ components of vector $\mathbf{B}$.

## 4. Reduction to Acín Canonical Form.

Our next step is to apply local transformations to maximally reduce vectors $\mathbf{A}$, $\mathbf{B}$, $\mathbf{C}$. The goal is to simplify the state to such an extent that its canonical Acín form [40] arise without calculating and diagonalizing Acín matrices $\mathbf{T}_{0,1}$ and $\mathbf{T}'_{0,1}$.

Local transformations do not affect entanglement, but they are modifying the set of $\mathbf{ABC}$ vectors. First, we will simplify vector $\mathbf{B}$. As we discussed above, first component of vector $\mathbf{B}$ vanishes for parameter $\chi = 1/2 \arctan(\Delta_{in} / \Omega_{in})$, equation (88). For normalized state $x^2 + y^2 = 1/2$ second component of vector $\mathbf{B}$ is equal to $1/2$. Since $x^2 - y^2 = 0$ and $x^2 + y^2 = 1/2$, we have $x^2 = y^2 = 1/4$. Thus quaternions $2\mathbf{x}$ and $2\mathbf{y}$ are unit quaternions which can be represented as unitary SU(2) operators. Therefore operator $(2\mathbf{y})^{-1}$ can be applied for local rotation on qubit $a$ or $c$ resulting in reduction of quaternion $\mathbf{y}$ to $1/2\,\mathbf{y}^{-1}\mathbf{y} \equiv 1/2$, thereby vector part reduces to $\vec{y} = 0$. After these transformations we have simple equations for $\mathbf{ABC}$ become.

$$\mathbf{A} = \vec{x}^T, \ \mathbf{B} = \{0, 1/2, ix_0\}, \ \mathbf{C} = \vec{x} \tag{96}$$

Next step is simplification of $\mathbf{AC}$. First observation: application of (81) and (92) with $\xi^c = (\xi^a)^{-1}$ is an adjoint action on $\mathbf{x}_{out} = (\xi^a)^{-1} \mathbf{x}_{out} \xi^a$ which evidently does not affect component $x_0$ while generating standard quaternionic representation of SO(3) rotation of vector $\vec{x}$. Therefore using basic algebra one can rotate both vectors $\mathbf{AC}$ to a simple form of one-component vectors, for example, $\mathbf{A} = \mathbf{C} = (0, 0, x_3)$. Since $x_0^2 + x_3^2 = 1/4$ it is convenient to use trigonometric parametrization $x_0 = 1/2 \sin\xi$ $x_3 = 1/2 \cos\xi$ such that (96) reads as follows

$$\mathbf{A} = \mathbf{C} = (0, 0, \cos\xi)/2, \ \mathbf{B} = (0, 1, i\sin\xi)/2 \tag{97}$$

Next step is to calculate $\mathbf{ABC}$ for Acín state and compare coefficients. We take Acín state in the $c(ab)$ form adjusted by a global phase

$$\left| \Psi^{Acin} \right\rangle = e^{i\pi/4} \left( \lambda_0 \left| 000 \right\rangle + \lambda_1 \left| 010 \right\rangle + \lambda_2 \left| 110 \right\rangle + \lambda_3 \left| 011 \right\rangle + \lambda_4 \left| 111 \right\rangle \right) \tag{98}$$

As a result we have from (3a) and (3c)

$$\mathbf{A} = (\lambda_0\lambda_3, i\lambda_0\lambda_3, -\lambda_0\lambda_4)$$
$$\mathbf{C} = (\lambda_0\lambda_2, i\lambda_0\lambda_2, -\lambda_0\lambda_4) \tag{99}$$
$$\mathbf{B} = (\lambda_2\lambda_3 - \lambda_1\lambda_4, i\lambda_1\lambda_4 - i\lambda_2\lambda_3, -\lambda_0\lambda_4)$$

Comparing (97) and (99) one can identify the apropriate combination of parameters $\lambda$. First, imaginary parts of $\mathbf{AC}$ vanish for $\lambda_2 = 0$, $\lambda_3 = 0$. Next we have $\lambda_0\lambda_4 = -\cos\xi$, $\lambda_1\lambda_4 = \sin\xi$ such that (99) becomes

$$\mathbf{A} = \mathbf{C} = (0, 0, \cos\xi)$$
$$\mathbf{B} = (-\sin\xi, i\sin\xi, \cos\xi) \tag{100}$$

To satisfy normalization $\lambda_0^2 + \lambda_1^2 + \lambda_4^2 = 1$ we take $\lambda_4 = 1/\sqrt{2}$, $\lambda_0 = -1/\sqrt{2}\cos\xi$, $\lambda_1 = 1/\sqrt{2}\sin\xi$, such that state (98) becomes

$$\left|\Psi^{Acin}\right\rangle = e^{i\pi/4}\left(-\cos\xi\left|000\right\rangle + \sin\xi\left|010\right\rangle + \left|111\right\rangle\right)/\sqrt{2} \tag{101}$$

This equation indicates that quaternionic states belong to a subset of three-parameter Acín class of locally equivalent states.

Finally, $\pi/2$ rotation in 2-3 plane and subsequent rotation by angle $\xi$ in 1-2 plane on qubit $b$ transform vector $\mathbf{B}$ in equation (100) to match $\mathbf{B}$ in equation (97).

## XV. SUMMARY

We have introduced vectors $\mathbf{A}$, $\mathbf{B}$ and $\mathbf{C}$ as invariants of two-qubit local $SL(2)^{(b,c)}$, $SL(2)^{(a,c)}$, and $SL(2)^{(a,b)}$ transformations. These vectors obey Bloch-type $SO(3)$ evolution under local operations acting on qubits $a$, $b$, $c$ correspondingly. These vectors define three-tangle and two-qubit concurrences. We provided $SO(6)$ representation of qubit-qubit $SU(4)$ coupling operators, and as an examples of $SO(6)$ form of $SU(4)$ coupling we considered CNOT, CZ, and SWAP gates. We showed how to use 6-D geometry to construct an arbitrary two-qubit transformation by concatenating three CZ-type gates. We applied $SO(6)$ geometry to find an efficient control sequence for transformation of asymmetric $\mathbf{W}$-type states to $\mathbf{GHZ}$ state. We have described geometric properties of three-tangle and provided examples of $SO(6)$ transformations maximizing three-tangle. Finally, we discussed entanglement control by $USp(4)$ quaternionic subgroup of $SU(4)$ group.

## APPENDIX: ISOMORPHYSM BETWEEN su(4) AND so(6) LIE ALGEBRAS

Formally, isomorphism between $su(4)$ and $so(6)$ algebras follows from the fact that $A_3$ and $D_3$ Dynkin diagrams are identical (see pages 150, 161 of ref. [8] or chapter 10 of ref. [10]). In our case, specific realization of Lie algebra isomorphism between $su(4)$ and $so(6)$ algebras is calculated explicitly by analyzing infinitesimal action of all fifteen $su(4)^{(ab)}$ generators on the space spanned by polynomials $\mathbf{A}$ and $\mathbf{B}$ defined by equations (3abc). One can equivalently choose pairs $\mathbf{A}$ and $\mathbf{C}$ or $\mathbf{B}$ and $\mathbf{C}$. Note that $\mathbf{A} \oplus \mathbf{B}$ space is an invariant space of $SU(4)^{(a,b)}$ action on $c_{ijk}$ coefficients in equation (1). As a result one obtains the following map between generators (see section V ref. [1] for details).

$$i/2 \begin{pmatrix} \sigma_z^{(a)} & & & & \\ \sigma_y^{(a)} & \sigma_x^{(a)} & & & \\ \sigma_{xx}^{(ab)} & \sigma_{xy}^{(ab)} & \sigma_{xz}^{(ab)} & & \\ \sigma_{yx}^{(ab)} & \sigma_{yy}^{(ab)} & \sigma_{yz}^{(ab)} & \sigma_z^{(b)} & \\ \sigma_{zx}^{(ab)} & \sigma_{zy}^{(ab)} & \sigma_{zz}^{(ab)} & \sigma_y^{(b)} & \sigma_x^{(b)} \end{pmatrix} \rightarrow \begin{pmatrix} l_{2,1}^{(a)} & & & & \\ -l_{3,1}^{(a)} & l_{3,2}^{(a)} & & & \\ \Lambda_{1,1}^{(ab)} & \Lambda_{1,2}^{(ab)} & \Lambda_{1,3}^{(ab)} & & \\ \Lambda_{2,1}^{(ab)} & \Lambda_{2,2}^{(ab)} & \Lambda_{2,3}^{(ab)} & l_{2,1}^{(b)} & \\ \Lambda_{3,1}^{(ab)} & \Lambda_{3,2}^{(ab)} & \Lambda_{3,3}^{(ab)} & -l_{3,1}^{(b)} & l_{3,2}^{(b)} \end{pmatrix} \tag{A1}$$

Matrices $\Lambda_{3,1}^{(ab)}$ are given by equation (18). Operators $l_{n,m}^{(a)}$ and $l_{n,m}^{(b)}$ are local rotations, i.e. 3x3 matrices defined in equation (6a) embedded as block-diagonal matrices in a 6x6 matrix acting on $\mathbb{R}^6$ space $\mathbf{A} \oplus \mathbf{B}$.

$$\left[l_{n,m}^{(a)}\right]_{i,j} = -\delta_{i,n}\delta_{j,m} + \delta_{i,m}\delta_{j,n}, \quad i,j = 1,2...6$$
$$\left[l_{n,m}^{(b)}\right]_{i,j} = -\delta_{i,n+3}\delta_{j,m+3} + \delta_{i,m+3}\delta_{j,n+3}, \quad i,j = 1,2...6$$
(A2)

Map (A1) has been reproduced in ref. [1] (Appendix: SU(4)-SO(6) Homomorphism) using Clifford algebra method for constructing generators of Spin groups [12], i.e. double covers of SO groups. This map preserves all commutation relations between 15 generators usually represented as 15x15 table of 225 commutators (e.g. Table I of ref. [41]). One can directly verify that commutation relations are preserved. Here are examples of typical commutation relations.

$$[i/2\sigma_{x,y,z}^{(a)}, i/2\sigma_{x,y,z}^{(b)}] = 0 \quad \Leftrightarrow \quad [l_{i,j}^{(a)}, l_{n,m}^{(a)}] = 0$$
$$[i/2\sigma_x^{(a)}, i/2\sigma_{x(x,y,z)}^{(ab)}] = 0 \quad \Leftrightarrow \quad [l_{3,2}^{(a)}, \Lambda_{1,(1,2,3)}^{(ab)}] = 0$$
$$[i/2\sigma_y^{(a)}, i/2\sigma_{x(x,y,z)}^{(ab)}] = i/2\sigma_{z(x,y,z)}^{(ab)} \quad \Leftrightarrow \quad [(-l_{3,1}^{(a)}), \Lambda_{1,(1,2,3)}^{(ab)}] = \Lambda_{3,(1,2,3)}^{(ab)}$$
$$[i/2\sigma_{y(xyz)}^{(ab)}, i/2\sigma_{x(xyz)}^{(ab)}] = i/2\sigma_z^{(a)} \quad \Leftrightarrow \quad [\Lambda_{2,(1,2,3)}^{(ab)}, \Lambda_{1,(1,2,3)}^{(ab)}] = l_{2,1}^{(a)}$$
$$[i/2\sigma_{yx}^{(ab)}, i/2\sigma_{xz}^{(ab)}] = 0 \quad \Leftrightarrow \quad [\Lambda_{2,1}^{(ab)}, \Lambda_{1,3}^{(ab)}] = 0$$

## ACKNOWLEDGMENTS

P.M.A. would like to acknowledge support of this work from the Air Force Office of Scientific Research (AFOSR). D.B.U. would like to acknowledge support for this work provided by the AFOSR Summer Faculty Fellowship Program under Grant No. FA8750-20-3-1003. Any opinions, findings, and conclusions or recommendations expressed in this material are those of the author(s) and do not necessarily reflect the views of the Air Force Research Laboratory.